\documentclass[prb,tighten]{revtex4-1}

\usepackage{amssymb,amsmath}
\newcommand{\dd}{\text{d}}
\usepackage{color}

\usepackage{graphicx}

\begin{document}
\title{Diffusion transport coefficients for granular binary mixtures at low density. Thermal diffusion segregation}
\author{Vicente Garz\'{o}}
\email{vicenteg@unex.es}
\homepage{URL: http://www.unex.es/eweb/fisteor/vicente/}
\affiliation{Departamento de F\'{\i}sica, Universidad de Extremadura, E-06071 Badajoz, Spain}
\author{J. Aaron Murray}
\affiliation{Department of Chemical and Biological Engineering,
University of Colorado, Boulder, CO 80309}
\author{Francisco Vega Reyes}
\email{fvega@unex.es}
\homepage{URL: http://www.unex.es/eweb/fisteor/fran/}
\affiliation{Departamento de F\'{\i}{s}ica, Universidad de Extremadura, E-06071 Badajoz, Spain}

\begin{abstract}
The mass flux of a low-density granular binary mixture obtained previously by solving the Boltzmann equation by means of the Chapman-Enskog method is considered further. As in the elastic case, the associated transport coefficients $D$, $D_p$ and $D'$ are given in terms of the solutions of a set of coupled linear integral equations which are approximately solved by considering the first and second Sonine approximations. The diffusion coefficients are explicitly obtained as functions of the coefficients of restitution and the parameters of the mixture (masses, diameters and concentration) and their expressions hold for an arbitrary number of dimensions. In order to check the accuracy of the second Sonine correction for highly inelastic collisions, the Boltzmann equation is also numerically solved by means of the direct simulation Monte Carlo (DSMC) method to determine the mutual diffusion coefficient $D$ in some special situations (self-diffusion problem and tracer limit).
The comparison with DSMC results reveals that the second Sonine approximation
to $D$ improves the predictions made from the first Sonine approximation. 
We also study the granular segregation driven by a uni-directional thermal gradient.
The segregation criterion is obtained from the so-called thermal diffusion factor $\Lambda$, which measures the amount of segregation parallel to the temperature gradient. The factor $\Lambda$ is determined here
by considering the second-order Sonine forms of the diffusion coefficients and its dependence on the coefficients of restitution is widely analyzed across the parameter space of the system. The results obtained in this paper extend previous works carried out in the tracer limit (vanishing mole fraction of one of the species) by some of the authors of the present paper.
\end{abstract}

\draft

\date{\today}
\maketitle

\section{Introduction}
\label{sec1}

It is well established that granular matter under rapid flow conditions admits a hydrodynamic-like-description. At sufficiently low density, the Boltzmann kinetic equation conveniently adapted to account for the inelastic character of collisions \cite{GS95,BDS97,BP04} has been used as the starting point to derive the corresponding hydrodynamic equations. The essential assumption to get those equations is the existence of a \emph{normal} solution, \cite{CC70} defined to be one for which all the space and time dependence occurs through a functional dependence on the hydrodynamic fields. In the case of small spatial gradients, the Chapman-Enskog method \cite{CC70} provides a constructive means to get this normal solution and in particular, to obtain the Navier-Stokes (NS) constitutive equations in the first order of the expansion. In this context, the study of hydrodynamics for granular gases follows similar steps as those made for ordinary gases.

On the other hand, as in the elastic case, \cite{CC70} the explicit form of the corresponding NS transport coefficients requires the solution of a set of linear integral equations. The standard procedure of solving these integral equations consists of expanding the solutions in Sonine polynomials. \cite{CC70} For simplicity, usually only the lowest Sonine polynomial (first Sonine approximation) is retained. However, in spite of this simple approximation, the results obtained from this approach compare in general well with Monte Carlo simulations \cite{BRC99,GM02,BR04} for mild degrees of inelasticity. Although most of the theoretical results \cite{BDKS98,GD99,L05,GSM07,NBSG07} have been devoted to monocomponent gases, some progresses have been made in the past few years in the case of granular mixtures (namely, systems composed by grains of different masses, diameters and concentrations). In particular, in the context of granular mixtures at low density, Garz\'o and Dufty \cite{GD02} have developed a kinetic theory which covers some aspects not completely covered in previous works. \cite{JM89,Z95,AW98,WA99,SGNT06,Serero09} Specifically, (i) the Garz\'o-Dufty theory goes beyond the weak dissipation limit so that it is expected to be applicable to a wide range of coefficients of restitution, and (ii) it takes into account the effects of nonequipartition of granular energy on the NS transport coefficients. As in the case of simple granular gases, the accuracy of the predictions of the Garz\'o-Dufty theory (which are based on the first Sonine approximation) has been confirmed by numerical solutions of the (inelastic) Boltzmann equation by means of the direct simulation Monte Carlo (DSMC) method \cite{B94} in the cases of the tracer diffusion coefficient \cite{GM04,GM07} and the shear viscosity coefficient of a driven mixture. \cite{MG03a,GM07} However, and contrary to the monocomponent case, discrepancies between theory and simulation appear to be important at strong dissipation for disparate mass and/or disparate size binary mixtures. Recently, the Garz\'o-Dufty theory has been extended to moderately dense binary mixtures \cite{GDH07} and the theoretical predictions compare also quite well with computer simulations. \cite{GM03b,GF09}

A possible way of reducing the discrepancies between theory and DSMC results is to consider higher-order terms in the Sonine polynomial expansion. In fact, recent works \cite{GF09} analyzing diffusion of impurities in a granular gas have shown that the second Sonine approximation to the tracer diffusion coefficient yields a dramatic improvement (up to 50║\%) over the first Sonine approximation when impurities are lighter than the surrounding gas in the range of large inelasticity. The results also show that the differences between the second Sonine approach and computer simulations are in general small (less than 4\%) for arbitrarily large inelasticity. This good agreement stimulates the evaluation of the complete set of NS transport coefficients of a granular binary mixture (with \emph{arbitrary} relative concentration) by retaining terms up to the second Sonine approximation. On the other hand, needless to say, the above goal is quite intricate due to the large number of collision integrals involved in the calculation. In this paper, we will cover partially this ambitious project by addressing the evaluation of the transport coefficients associated with the mass flux.

We consider a binary mixture composed by smooth inelastic disks
($d=2$) or spheres ($d=3$) of masses $m_{1}$ and $ m_{2} $, and
diameters $\sigma _{1}$ and $\sigma _{2}$. The inelasticity of
collisions among all pairs is characterized by three independent
constant coefficients of restitution\cite{GD02} $\alpha _{11}$, $\alpha
_{22}$, and $\alpha _{12}=\alpha _{21}$, where $\alpha _{ij}\leq
1$ is the coefficient of restitution for collisions between
particles of species $i$ and $j$. The case $\alpha_{ij}=1$ corresponds to elastic collisions. To first order in the spatial gradients, the constitutive equation for the mass flux $\textbf{j}_i$ (with $i=1,2$) is given by \cite{GD02}
\begin{equation}
\label{1.1}
\mathbf {j}_1=-\frac{m_1m_2n}{\rho}D\nabla x_1-\frac{\rho}{p}D_p \nabla p
-\frac{\rho}{T}D' \nabla T, \quad \mathbf{j}_2=-\mathbf{j}_1,
\end{equation}
where $D$ is the (mutual) diffusion coefficient, $D_p$ is the pressure diffusion coefficient, and $D'$ is the thermal diffusion coefficient. Here, $n=n_1+n_2$ is the total number density ($n_i$ is the number density of species $i$), $\rho=m_1n_1+m_2n_2$ is the total mass density, $x_i=n_i/n$ is the concentration (or mole fraction) of species $i$, $T$ is the granular temperature of the mixture and $p=nT$ is the hydrostatic pressure. One of the goals of this paper is to determine the diffusion coefficients $D$, $D_p$, and $D'$ of a dilute granular mixture in terms of the coefficients of restitution $\alpha_{11}$, $\alpha_{22}$, and $\alpha_{12}$ and the parameters of the mixture (relative masses, diameters and concentration). As said before and in contrast to our previous works, \cite{GD02,GM07,GMD06} the diffusion coefficients will be explicitly obtained by considering contributions up to the second Sonine approximation.

There are several reasons to address the above calculation. First, given that the results reported in Refs.\ \onlinecite{GF09} are limited to the tracer limit ($x_1\to 0$), the question arises then as to whether (and if so, to what extent) the conclusions drawn before \cite{GM04,GF09} may apply when one considers arbitrary concentrations. This goal is not only academic since, from a practical standpoint, many computer simulations \cite{MG02,simulations} and experiments \cite{exp} in flowing granular mixtures involve \emph{finite} concentrations. As a second reason, it must be noted that previous results \cite{KCM87} obtained for ordinary mixtures (i.e.,
when the collisions are elastic)  have clearly shown that while the first Sonine approximation
can accurately describe the shear viscosity and the thermal conductivity coefficients,  it cannot
achieve the same degree of accuracy for the mutual and thermal diffusion coefficients.
In this latter case, Kincaid \emph{et al.} \cite{KCM87} concluded that the second Sonine approximation
is much better approximation than the first one for a wide range of values of masses and sizes.
A third motivation to improve the evaluation of the NS transport coefficients lies in the fact
that the reference homogeneous cooling state (HCS) is known to be unstable against long wavelength
spatial perturbations, leading to vortex and cluster formation. Since this instability can be well
characterized \cite{Peter11} through a linear stability analysis of the hydrodynamic equations, \cite{G05}
a more accurate evaluation of the NS transport coefficients for large inelasticity may help to understand
the physical mechanisms involved in this instability. Finally, as a fourth motivation and given that the
second Sonine approach is expected to differ from the first one at strong dissipation, the results reported
here can be of practical interest since the range of high inelasticities has growing interest
in experimental works \cite{exp1} and is also exhibited by wetted particles. \cite{DHDNZ10}

Since the explicit second-Sonine order expressions of $D$, $D_p$ and $D'$ are at hand, a segregation criterion based on thermal diffusion is derived. This is the second objective of the paper. Thermal diffusion is caused by the relative motion of the components of a mixture due to the presence of a temperature gradient. Under these conditions, a steady state can be reached in which the separation effect arising from thermal diffusion is balanced by the remixing effect of ordinary diffusion. As a consequence, segregation is observed and characterized by the so-called thermal diffusion factor $\Lambda$. While the factor $\Lambda$ has been previously studied \cite{KCM87} in ordinary mixtures by using the second-Sonine approximation, much less is known about thermal diffusion in granular mixtures. The present analysis complements previous studies \cite{GF09} carried out in the tracer limit by considering the second-Sonine order solution to the diffusion coefficients. As expected, the present results show that the effect of inelasticity of collisions on $\Lambda$ is in general quite significant.

An important issue that may lead to confusion is the applicability of the expression for the mass flux derived here in the first-order of the spatial gradients (NS hydrodynamic order). The forms of the three diffusion coefficients do not limit their application to weak inelasticity and hold in principle for arbitrary values of the coefficients of restitution. In fact, the results reported below include a domain of both weak and strong inelasticity, $0.5\leq \alpha_{ij}\leq 1$. On the other hand, as already pointed out in previous works, \cite{GMD06,GFM09} the NS hydrodynamic equations themselves may or may not be limited with respect to inelasticity, depending on the particular granular flow considered. While in the case of ordinary fluids the strength of the spatial gradients is controlled solely by the initial or boundary conditions, for granular gases the steady state conditions are controlled both by the boundary conditions and the degree of inelasticity in the collisions. \cite{G03,SGD04,VU09} An illustrative example of this coupling is the so-called LTu flow class, \cite{VSG10,VSG13} of which the well-known (steady) simple shear flow \cite{G03,SGD04} is a special case.  The LTu flow class (and thus, the simple shear flow) can only occur when there is an exact balance between the collisional cooling (which is fixed by the mechanical properties of the particles making up the granular fluid) and the viscous heating (which is essentially fixed by the shear rate). Unfortunately, except for the quasi-elastic limit ($\alpha_{ij} \simeq 1$), this balance only occurs for high shear rates and so, one needs to include higher order corrections (such as Burnett-order terms) to the NS solution. \cite{GT96,SG98} Consequently, the NS hydrodynamics would only be expected to work in \emph{steady} granular flows in the quasielastic limit. \cite{SGD04}

In spite of the above cautions, the NS description is still accurate and appropriate for a wide class of flows. One of them corresponds to small spatial perturbations of the HCS for an isolated system. Both molecular dynamics \cite{Peter11} and Monte Carlo simulations\cite{BRC99,Brey2} have confirmed the dependence of the NS transport coefficients on inelasticity (even in highly dissipative granular gases) and the reliability of the NS hydrodynamics to describe shearing instabilities. In the case of dense gases, the predictions of the Enskog kinetic theory \cite{GD99,L05} show both qualitative  and quantitative agreement with computer simulations \cite{L02,LLC07,BGGL09} and with real experiments of supersonic flow past a wedge (where there is no reason \emph{a priori} to expect that the NS approximation works well) \cite{R02} and nuclear magnetic experiments of a system of mustard seeds vibrated vertically. \cite{Y02} Therefore, the NS equations can still be considered as an important and useful tool to describe granular flows although more limited than for ordinary gases.

The plan of the paper is as follows. First, in Sec.\ \ref{sec2} the Boltzmann equation and its corresponding balance hydrodynamic equations for the mass, momentum and energy are recalled. In Sec.\ \ref{sec3}, the diffusion transport coefficients $D$, $D_p$, and $D'$ are given in terms of the solution of a set of coupled linear integral equations previously derived by Garz\'o and Dufty. \cite{GD02} These integral equations are \emph{approximately} solved by using the first and second Sonine approximations, where explicit forms for the above transport coefficients are provided. Technical details of the calculations carried out in this paper are relegated to two Appendices. Next, the theoretical approaches (first and second Sonine approximations) are compared in Sec.\ \ref{sec4} with available and new simulation data obtained from numerical solutions of the Boltzmann equation by using the DSMC method for the self-diffusion and tracer diffusion coefficients. Two- and three-dimensional systems are considered. The dependence of the complete set of diffusion coefficients on the parameter space of the system is widely analyzed in Sec.\ \ref{sec5}. The results indicate that, while the convergence of the Sonine expansion on the mutual diffusion coefficient $D$ is relatively good, in the cases of the coefficients $D_p$ and $D'$ is not as good. Segregation by thermal diffusion is studied in Sec.\ \ref{sec6} and  the paper is closed in Sec.\ \ref{sec7} with a brief discussion of the results.

\section{Boltzmann kinetic theory for granular binary mixtures. Chapman-Enskog method}
\label{sec2}

Let us consider a granular binary mixture where the density of each species is sufficiently low. In this case, all the relevant information on the state of the mixture is described by the velocity distribution functions $f_{i}({\bf r},{\bf v};t)$  of each species $(i=1,2)$. These distributions obey the set of nonlinear Boltzmann equations \cite{BDS97}
\begin{equation}
\left( \partial _{t}+{\bf v}\cdot\nabla \right) f_{1}({\bf r},{\bf
v},t)=J_{11}\left[ {\bf v}|f_{1}(t),f_{1}(t)\right]+J_{12}\left[ {\bf v}|f_{1}(t),f_{2}(t)\right], \label{2.1}
\end{equation}
and an analogous equation for $f_{2}({\bf r},{\bf v};t)$. The Boltzmann collision operators $J_{ij}\left[ {\bf
v}|f_{i},f_{j}\right] $ are given by
\begin{equation}
J_{ij}\left[ {\bf v}_{1}|f_{i},f_{j}\right] =\sigma
_{ij}^{d-1}\int \dd{\bf v} _{2}\int \dd\widehat{\boldsymbol {\sigma
}}\,\Theta (\widehat{{\boldsymbol {\sigma }}} \cdot {\bf
g}_{12})(\widehat{\boldsymbol {\sigma }}\cdot {\bf g}_{12})
\left[ \alpha _{ij}^{-2}f_{i}({\bf r},{\bf v}_{1}^{\prime
},t)f_{j}( {\bf r},{\bf v}_{2}^{\prime },t)
-f_{i}({\bf r},{\bf v}
_{1},t)f_{j}({\bf r}, {\bf v}_{2},t)\right], \label{2.2}
\end{equation}
where $d$ is the dimensionality of the system,
$\sigma _{ij}=\left( \sigma _{i}+\sigma _{j}\right) /2$,
$\widehat{\boldsymbol {\sigma}}$ is a unit vector along the line
of centers, $\Theta $ is the Heaviside step function, and ${\bf
g}_{12}={\bf v}_{1}-{\bf v}_{2}$ is the relative velocity. The
primes on the velocities denote the initial values $\{{\bf
v}_{1}^{\prime }, {\bf v}_{2}^{\prime }\}$ that lead to $\{{\bf
v}_{1},{\bf v}_{2}\}$ following a binary (restituting) collision:
\begin{equation}
{\bf v}_{1}^{\prime }={\bf v}_{1}-\mu _{ji}\left( 1+\alpha
_{ij}^{-1}\right) (\widehat{{\boldsymbol {\sigma }}}\cdot {\bf
g}_{12})\widehat{{\boldsymbol {\sigma }}} ,\nonumber\\
\end{equation}
\begin{equation}
 {\bf v}_{2}^{\prime }={\bf v}_{2}+\mu _{ij}\left( 1+\alpha
_{ij}^{-1}\right) (\widehat{{\boldsymbol {\sigma }}}\cdot {\bf
g}_{12})\widehat{ \boldsymbol {\sigma}},  \label{2.3}
\end{equation}
where $\mu _{ij}\equiv m_{i}/\left( m_{i}+m_{j}\right) $.

In the case of granular mixtures, the relevant hydrodynamic fields are the number densities $n_{i}(\textbf{r},t)$, the flow velocity $\textbf{u}(\textbf{r},t)$, and the granular temperature $T(\textbf{r},t)$. In terms of the velocity distribution functions $f_{i}({\bf r},{\bf v},t)$, the above fields are defined respectively as
\begin{equation}
n_{i}=\int \dd{\bf v}f_{i}({\bf v})\;,\quad \rho {\bf u}=\sum_{i=1}^2m_{i}
\int \dd {\bf v}{\bf v}f_{i}({\bf v}),
\label{2.4}
\end{equation}
\begin{equation}
T=\sum_{i=1}^2 x_i T_i=\sum_{i=1}^2\frac{m_{i}}{d n}\int \dd{\bf
v}V^{2}f_{i}({\bf v}), \label{2.5}
\end{equation}
where $\rho=m_1n_1+m_2n_2$ is the total mass density and ${\bf V}={\bf v}-{\bf u}$ is the peculiar velocity. The third equality of Eq.\ (\ref{2.5})
defines the kinetic temperatures $T_i$ for each species, which
measure their mean kinetic energies. The exact macroscopic balance equations for $n_{i}(\textbf{r},t)$, $\textbf{u}(\textbf{r},t)$, and $T(\textbf{r},t)$ follow directly from Eq.\ \eqref{2.1} (and its corresponding counterpart for $f_2$) by multiplying with $1$, $m_i \textbf{v}$, and $\frac{1}{2}m_iv^2$ and integrating over $\textbf{v}$. They are given by \cite{GD02}
\begin{equation}
D_{t}n_{i}+n_{i}\nabla \cdot {\bf u}+\frac{\nabla \cdot {\bf
j}_{i}}{m_{i}} =0,  \label{2.6}
\end{equation}
\begin{equation}
D_{t}{\bf u}+\rho ^{-1}\nabla \cdot {\sf P}={\bf 0},
\label{2.7}
\end{equation}
\begin{equation}
D_{t}T-\frac{T}{n}\sum_{i=1}^2\frac{\nabla \cdot {\bf
j}_{i}}{m_{i}}+\frac{2}{dn} \left( \nabla \cdot {\bf q}+{\mathsf
P}:\nabla {\bf u}\right) =-\zeta \,T. \label{2.8}
\end{equation}
In the above equations, $D_{t}=\partial _{t}+{\bf u}\cdot \nabla $
is the material derivative,
\begin{equation}
{\bf j}_{i}=m_{i}\int \dd{\bf v}\,{\bf V}\,f_{i}({\bf v})
\label{2.9}
\end{equation}
is the mass flux for species $i$ relative to the local flow,
\begin{equation}
\mathsf{P}=\sum_{i=1}^2\,m_i\,\int \dd{\bf v}\,{\bf V}{\bf
V}\,f_{i}({\bf v})  \label{2.10}
\end{equation}
is the total pressure tensor,
\begin{equation}
{\bf q}=\sum_{i=1}^2\,\frac{m_i}{2}\int \dd{\bf v}\,V^{2}{\bf
V}\,f_{i}({\bf v})  \label{2.11}
\end{equation}
is the total heat flux, and
\begin{equation}
\zeta=\sum_{i=1}^2\;x_i \gamma_i \zeta_i=
-\frac{1}{p}\sum_{i=1}^2\sum_{j=1}^2\frac{m_i}{d}\int \dd{\bf v}V^{2}J_{ij}[{\bf v}
|f_{i},f_{j}],  \label{2.12}
\end{equation}
is the total ``cooling rate'' due to inelastic collisions among all species. In Eq.\ \eqref{2.12}, $p=nT=\frac{1}{d}\text{Tr}\mathsf {P}$ is the hydrostatic pressure, $\gamma_i\equiv T_i/T$ and the second equality defines the ``cooling rates'' $\zeta_i$ for the partial temperatures $T_i$. \cite{GD02}

The balance equations \eqref{2.6}--\eqref{2.8} do not constitute a closed set of equations for the hydrodynamic fields unless one knows the functional dependence of $\textbf{j}_i$, $\mathsf{P}$, $\textbf{q}$, and $\zeta$ on the above fields. On the other hand, for times longer than the mean free time, the distribution functions $f_i$ are expected to adopt the form of a \emph{normal} or hydrodynamic solution such that all space and time dependence of $f_i$ occurs through the hydrodynamic fields:
\begin{equation}
f_i({\bf r},{\bf v},t)=f\left[{\bf v}|x_1 ({\bf r}, t), p({\bf r}, t),
T({\bf r}, t), {\bf u}({\bf r}, t) \right].
\label{2.13}
\end{equation}
Note that we have taken the set $\{x_1, p, T, {\bf u}\}$ as the $d+3$ independent fields of the
two-component mixture. As mentioned in Ref.\ \onlinecite{GMD06}, in the case of inelastic systems, there is more flexibility than in ordinary mixtures to chose the set of relevant hydrodynamic fields since the specific set of gradients contributing to each flux is only restricted by fluid symmetry considerations. Here, as in our previous works for dilute granular mixtures, \cite{GD02,GMD06} we have chosen the set $\{x_1, p, T, {\bf u}\}$ since they are the most accessible fields from an experimental point of view. In particular, a contribution proportional to $\nabla p$ (which is absent in the elastic case) appears in the mass and heat fluxes.

In the case of small spatial variations (i.e., low Knudsen numbers), the functional dependence \eqref{2.13} can be made local in space through an expansion in the gradients of the hydrodynamic fields. This is the procedure followed in the Chapman-Enskog method \cite{CC70} to get an approximate solution to the Boltzmann equation. Thus, the distributions $f_i$ are written as
\begin{equation}
f_i=f_i^{(0)}+\epsilon \,f_i^{(1)}+\epsilon^2
\,f_i^{(2)}+\cdots, \label{2.14}
\end{equation}
where each factor of $\epsilon$ (formal non-uniformity parameter) means an implicit gradient of a
hydrodynamic field. In the first-order of the expansion, the NS constitutive equations for the mass, momentum and heat fluxes can be derived. In this paper, we will focus  our attention to the first-order contribution $\textbf{j}_i^{(1)}$ to the mass flux.

As said in the Introduction, in the case of ordinary gases ($\alpha_{ij}=1$) the strength of the spatial gradients is imposed by the boundary or initial conditions. However, the situation is more complicated for granular gases ($\alpha_{ij}\neq 1$) since for steady states \cite{G03,SGD04,VSG10} the size of the spatial gradients is set by boundary conditions and inelasticity together. Therefore, the NS equations are in principle expected to be reliable for steady granular flows just in the case of nearly elastic particles since inelasticity may set by itself large gradients. \cite{VU09} In the Chapman-Enskog solution worked out here, we have assumed that the spatial gradients are independent of the coefficients of restitution $\alpha_{ij}$ and so the corresponding diffusion transport coefficients hold for arbitrary values of $\alpha_{ij}$.\cite{GD02} It must remarked that our perturbation scheme differs from previous works on granular mixtures \cite{JM89,SGNT06} where the Chapman-Enskog solution is given in powers of both the spatial gradients (or equivalently, the Knudsen number) and the degree of dissipation $\xi_{ij}\equiv 1-\alpha_{ij}^2$. In fact, in those works\cite{JM89,SGNT06} the reference distribution functions $f_i^{(0)}$ are chosen to be Maxwellians at the same temperature ($T_1=T_2=T$), ignoring the real effect of energy non-equipartition in granular mixtures. \cite{GD02} As a consequence, the results provided in Refs.\ \onlinecite{JM89,SGNT06} only agree with our results in the quasielastic limit ($\xi_{ij}\simeq 0$).

\section{Diffusion transport coefficients}
\label{sec3}

The application of the Chapman-Enskog method to the Boltzmann equation allows one to determine the form of the
NS transport coefficients of the mixture. In particular, the mass flux $\textbf{j}_i^{(1)}$ is given by
Eq.\ \eqref{1.1} where the diffusion transport coefficients $D$, $D_p$, and $D'$ are defined, respectively, as
\begin{equation}
D=-\frac{\rho }{dm_{2}n}\int \dd{\bf v}\,{\bf V}{\bf \cdot \,}{\boldsymbol {\cal A}}
_{1},  \label{3.2}
\end{equation}
\begin{equation}
D_{p}=-\frac{m_{1}p}{d\rho}\int \dd{\bf v}\,{\bf V}\,\cdot {\boldsymbol {\cal B}}_{1},  \label{3.3}
\end{equation}
\begin{equation}
D'=-\frac{m_{1}T}{d\rho }\int \dd{\bf v}\,{\bf V}\,\cdot {\boldsymbol {\cal C}}_{1}.
\label{3.4}
\end{equation}
As in the case of elastic collisions, \cite{CC70,FK} the quantities ${\boldsymbol {\cal A}}_{i}$,
${\boldsymbol {\cal B}}_{i}$, and ${\boldsymbol {\cal C}}_{i}$ ($i=1,2$) are the solutions of the
following set of coupled linear integral equations: \cite{GD02}
\begin{equation}
\left[ -\zeta ^{(0)}\left( T\partial _{T}+p\partial _{p}\right)
+{\cal L}_{1} \right] {\boldsymbol {\cal A}}_{1}+{\cal
M}_{1}{\boldsymbol {\cal A}}_{2}=\mathbf{A}_{1}+\left( \frac{\partial
\zeta ^{(0)}}{\partial x_{1}}\right) _{p,T}\left( p{\boldsymbol
{\cal B}}_{1}+T{\boldsymbol {\cal C}}_{1}\right),  \label{3.5a}
\end{equation}
\begin{equation}
\left[ -\zeta ^{(0)}\left( T\partial _{T}+p\partial _{p}\right)
+{\cal L}_{2} \right] {\boldsymbol {\cal A}}_{2}+{\cal
M}_{2}{\boldsymbol {\cal A}}_{1}=\mathbf{A}_{2}+\left( \frac{\partial
\zeta ^{(0)}}{\partial x_{1}}\right) _{p,T}\left( p{\boldsymbol
{\cal B}}_{2}+T{\boldsymbol {\cal C}}_{2}\right),  \label{3.5b}
\end{equation}
\begin{equation}
\left[ -\zeta ^{(0)}\left( T\partial _{T}+p\partial _{p}\right)
+{\cal L}_{1}-2\zeta ^{(0)}\right] {\boldsymbol {\cal B}}_{1}+{\cal M}_{1}{\boldsymbol {\cal B}}_{2}=
\mathbf {B}_{1}
+\frac{T\zeta ^{(0)}}{p}{\boldsymbol {\cal C}}_{1},
\label{3.6a}
\end{equation}
\begin{equation}
\left[ -\zeta ^{(0)}\left( T\partial _{T}+p\partial _{p}\right)
+{\cal L} _{2}-2\zeta ^{(0)}\right] {\boldsymbol {\cal
B}}_{2}+{\cal M}_{2}{\boldsymbol {\cal B}}_{1}= {\bf
B}_{2}+\frac{T\zeta ^{(0)}}{p}{\boldsymbol {\cal C}}_{2},
\label{3.6b}
\end{equation}
\begin{equation}
\left[ -\zeta ^{(0)}\left( T\partial _{T}+p\partial _{p}\right)
+{\cal L}_{1}-\frac{1}{2}\zeta ^{(0)}\right] {\boldsymbol {\cal
C}}_{1}+{\cal M}_{1} {\boldsymbol {\cal C}}_{2}= \mathbf {C}_{1}-\frac{p\zeta ^{(0)}}{2T}{\boldsymbol {\cal B}}_{1},
\label{3.7a}
\end{equation}
\begin{equation}
\left[ -\zeta ^{(0)}\left( T\partial _{T}+p\partial _{p}\right)
+{\cal L}_{2}-\frac{1}{2}\zeta ^{(0)}\right] {\boldsymbol {\cal
C}}_{2}+{\cal M}_{2} {\boldsymbol {\cal C}}_{1}= {\bf C}_{2}
-\frac{p\zeta ^{(0)}}{2T}{\boldsymbol {\cal B}}_{2},
\label{3.7b}
\end{equation}
Here, $\zeta^{(0)}=\zeta_1^{(0)}=\zeta_2^{(0)}$ is the cooling rate evaluated with the zeroth-order distribution and we have introduced the linearized Boltzmann collision operators
\begin{equation}
{\cal L}_{1}X=-\left(
J_{11}[f_{1}^{(0)},X]+J_{11}[X,f_{1}^{(0)}]+
J_{12}[X,f_{2}^{(0)}]\right),
\label{3.9}
\end{equation}
\begin{equation}
{\cal M}_{1}X=-J_{12}[f_{1}^{(0)},X].  \label{3.10}
\end{equation}
The corresponding forms for the operators ${\cal L}_{2}$ and ${\cal M}_{2}$ can be easily obtained from Eqs.\ \eqref{3.9} and \eqref{3.10}, respectively, by just making the changes 1$\leftrightarrow$2. In addition,
\begin{equation}
{\bf A}_{i}({\bf V})=-\left(\frac{\partial}{\partial x_{1}}
f_{i}^{(0)}\right)_{p,T}{\bf V},  \label{3.11}
\end{equation}
\begin{equation}
{\bf B}_{i}({\bf V})=-\frac{1}{p}\left[ f_{i}^{(0)}{\bf V}+\frac{nT}{\rho }
\left(\frac{\partial}{\partial {\bf V}}f_{i}^{(0)}\right) \right] ,
\label{3.12}
\end{equation}
\begin{equation}
{\bf C}_{i}({\bf V})=\frac{1}{T}\left[
f_{i}^{(0)}+\frac{1}{2}\frac{\partial }{\partial {\bf V}}\cdot \left( {\bf V}
f_{i}^{(0)}\right) \right] {\bf V}. \label{3.13}
\end{equation}

It is worthwhile remarking that so far the expressions for the transport coefficients $D$, $D_p$, and $D'$ are \emph{exact}. However, in order to determine the dependence of the above coefficients on the parameters of the mixture, one needs to solve the integral equations \eqref{3.5a}--\eqref{3.7b} and to know the explicit form of the (local) HCS distributions $f_i^{(0)}$. With respect to this latter point, both theoretical \cite{GD99a} and computer simulation \cite{MG02,simulations} results have shown that in the region of thermal velocities $f_i^{(0)}(\textbf{V})$ is well represented by its Maxwellian form at the partial temperature $T_i$, i.e.,
\begin{equation}
\label{3.14}
f_i^{(0)}(\textbf{V}) \to f_{i,M}({\bf
V})= n_i\left(\frac{m_i}{2\pi T_i}\right)^{d/2}\exp\left(-
\frac{m_i V^2}{2T_i}\right).
\end{equation}
Thus, in order to get simple and accurate expressions for the diffusion transport coefficients, we will neglect here the non-Gaussian corrections to $f_i^{(0)}(\textbf{V})$. While these corrections are not important in the case of the mass flux and the pressure tensor, \cite{GMD06} the impact of them on the heat flux is not negligible in highly dissipative gases. \cite{BR04,GSM07} Accordingly, a theory incorporating the above non-Gaussian corrections does not seem in practice necessary for computing the diffusion transport coefficients.

Regarding the unknowns ${\boldsymbol {\cal A}}_{i}$, ${\boldsymbol {\cal B}}_{i}$, and ${\boldsymbol {\cal C}}_{i}$, the standard method consists of approximating them by Maxwellians (at different temperatures) times truncated Sonine polynomial expansions. For simplicity, usually only the lowest Sonine polynomial (first Sonine approximation) is retained \cite{GD99,GMD06,GM07} and the results obtained from this simple approach agree in general relatively well with numerical results \cite{BRCG00,MG03a} for granular mixtures obtained from the DSMC method. However, as for ordinary mixtures, \cite{KCM87} significant discrepancies between theory and simulation appear when one considers
disparate values of mass and diameter ratios at small values of the coefficients of restitution. We may expect that this disagreement could be mitigated in part if one considers higher-order terms in the Sonine polynomial expansion, much like in the case of the diffusion coefficient $D$ for the tracer limit ($x_1\to 0$). \cite{GM04,GF09} In particular, as said in the Introduction, it is shown that the accuracy of the second Sonine approximation for $D$ is much better than the first Sonine approximation when the tracer particles are lighter than the particles of the gas. Motivated by these results, our goal here is to evaluate the complete set of diffusion coefficients $D$, $D_p$, and $D'$ up to the second Sonine approximation as functions of the coefficients of restitution ($\alpha_{11}$, $\alpha_{22}$, and $\alpha_{12}$) and the parameters of the mixture (masses $m_i$, diameters $\sigma_i$ and concentration $x_1$). Therefore, the present analysis generalizes to \emph{arbitrary} concentration our previous theoretical results derived in the simple tracer limit case.

In the second Sonine approximation, the quantities ${\boldsymbol {\cal A}}_{i}$, ${\boldsymbol {\cal B}}_{i}$, and ${\boldsymbol {\cal C}}_{i}$ are approximated by
\begin{equation}
\label{3.15a}
{\boldsymbol {\cal A}}_{1}({\bf V})\to f_{1,M}\left[a_{1,1}{\bf V}+a_{1,2}{\bf S}_1({\bf V})
\right] ,
\end{equation}
\begin{equation}
\label{3.15b}
{\boldsymbol {\cal A}}_{2}({\bf V})\to f_{2,M}\left[a_{2,1}{\bf V}+a_{2,2}{\bf S}_2({\bf V})\right],
\end{equation}
\begin{equation}
\label{3.16a}
{\boldsymbol {\cal B}}_{1}({\bf V})\to f_{1,M}\left[b_{1,1}{\bf V}+b_{1,2}{\bf S}_1({\bf V})
\right] ,
\end{equation}
\begin{equation}
\label{3.16b}
{\boldsymbol {\cal B}}_{2}({\bf V})\to f_{2,M}\left[b_{2,1}{\bf V}+b_{2,2}{\bf S}_2({\bf V})\right],
\end{equation}
\begin{equation}
\label{3.17a}
{\boldsymbol {\cal C}}_{1}({\bf V})\to f_{1,M}\left[c_{1,1}{\bf V}+c_{1,2}{\bf S}_1({\bf V})
\right] ,
\end{equation}
\begin{equation}
\label{3.17b}
{\boldsymbol {\cal C}}_{2}({\bf V})\to f_{2,M}\left[c_{2,1}{\bf V}+c_{2,2}{\bf S}_2({\bf V})\right],
\end{equation}
where
\begin{equation}
\label{3.18}
{\bf S}_i({\bf V})=\left(\frac{1}{2}m_iV^2-\frac{d+2}{2}T_i\right){\bf V}.
\end{equation}
The coefficients $\{a_{i,1}, b_{i,1}, c_{i,1}\}$ are related to the transport coefficients $D$, $D_p$, and $D'$, respectively, as
\begin{equation}
a_{1,1}=-\frac{n_2T_2}{n_1T_1}a_{2,1}=-\frac{m_{1}m_{2}n}{\rho n_{1}T_{1}}D,
\label{3.19}
\end{equation}
\begin{equation}
b_{1,1}=-\frac{n_2T_2}{n_1T_1}b_{2,1}=-\frac{\rho }{pn_{1}T_{1}}D_{p},
\label{3.20}
\end{equation}
\begin{equation}
c_{1,1}=-\frac{n_2T_2}{n_1T_1}c_{2,1}=-\frac{\rho }{Tn_{1}T_{1}}D^{\prime}.
\label{3.21}
\end{equation}
Upon writing the first equalities in Eqs.\ (\ref{3.19})--(\ref{3.21}) use has been made of the property ${\bf j}_1^{(1)}=-{\bf j}_2^{(1)}$.  The coefficients $\{a_{i,2}, b_{i,2}, c_{i,2}\}$ are defined as
\begin{equation}
\left(
\begin{array}{c}
a_{i,2} \\
b_{i,2} \\
c_{i,2}
\end{array}
\right) =\frac{2}{d(d+2)}\frac{m_i}{n_iT_i^3}
\int d{\bf v}{\bf S}_i({\bf V})\cdot\left(
\begin{array}{c}
{\boldsymbol {\cal A}}_{i} \\
{\boldsymbol {\cal B}}_{i} \\
{\boldsymbol {\cal C}}_{i}
\end{array}
\right)
\label{3.22}
\end{equation}

The diffusion transport coefficients $D$, $D_p$, and $D'$ and the second Sonine coefficients $a_{i,2}$, $b_{i,2}$, and $c_{i,2}$ are determined by substitution of Eqs.\ \eqref{3.15a}--\eqref{3.17b} into the integral equations \eqref{3.5a}--\eqref{3.7b}, multiplication by $m_i \textbf{V}$ and $\textbf{S}_i(\textbf{V})$, and integration over velocity. The procedure is lengthy and follows similar mathematical steps as those made before \cite{GM04,GF09} in the tracer limit ($x_1 \to 0$). Technical details on this calculation have been relegated to the Appendix \ref{appA}.

For the sake of convenience, we introduce dimensionless forms for the diffusion coefficients as
\begin{equation}
\label{3.23}
D=\frac{\rho T}{m_{1}m_{2}\nu _{0}}D^*,\quad
D_{p}=\frac{nT}{\rho \nu _{0}}D_{p}^*,\quad D^{\prime
}=\frac{nT}{\rho \nu _{0}}D^{\prime*},
\end{equation}
where
\begin{equation}
\label{3.24}
\nu_0=\sqrt{\pi}n\sigma _{12}^{d-1}\sqrt{2T\frac{m_1+m_2}{m_1m_2}}
\end{equation}
is an effective collision frequency. According to the relations \eqref{3.19}--\eqref{3.21}, the (reduced) Sonine coefficients $a_{11}^*\equiv \nu_0 a_{11}$, $b_{11}^*\equiv p\nu_0 b_{11}$, and $c_{11}^*\equiv T \nu_0 c_{11}$ are given, respectively, as
\begin{equation}
\label{3.24.1}
a_{11}^*=-\frac{D^*}{x_1\gamma_1}, \quad b_{11}^*=-\frac{D_p^*}{x_1\gamma_1}, \quad
c_{11}^*=-\frac{D^{'*}}{x_1\gamma_1}.
\end{equation}
The three first elements of the column matrix
\begin{equation}
\label{3.24.2}
\mathsf{X}\equiv \{a_{1,1}^*; b_{1,1}^*; c_{1,1}^*; a_{1,2}^*; a_{2,2}^*; b_{1,2}^*; b_{2,2}^*; c_{1,2}^*; c_{2,2}^*\}
\end{equation}
provide the expressions of the second Sonine approximations $a_{11}^*[2]$, $b_{11}^*[2]$, and $c_{11}^*[2]$. In Eq.\ \eqref{3.24.2}, $a_{i,2}^*=T\nu_0 a_{i,2}$, $b_{i,2}^*=pT\nu_0 b_{i,2}$, and $c_{i,2}^*=T^2\nu_0 c_{i,2}$.
The matrix $\mathsf{X}$ is given by
\begin{equation}
\label{3.24.2}
\mathsf{X}=\mathsf{\Omega}^{-1}\cdot \mathsf{Y},
\end{equation}
where $\mathsf{\Omega}$ is the 9$\times$9 square matrix defined by Eq.\ \eqref{a28bis} while the column matrix $\mathsf{Y}$ is given by Eq.\ \eqref{a31}. Once the above Sonine coefficients are known, the forms of
the (reduced) second Sonine diffusion coefficients $D^*[2]$, $D_p^*[2]$, and $D^{'*}[2]$ can be easily derived from the relations \eqref{3.24.1}. The expressions of the diffusion coefficients are analytic for any dimension $d$ and give $D^*[2]$, $D_p^*[2]$, and $D^{'*}[2]$ as functions of the mole fraction $x_1$, the mass ratio $\mu\equiv m_1/m_2$, the diameter ratio $\omega\equiv \sigma_1/\sigma_2$, and the coefficients of restitution $\alpha_{11}$, $\alpha_{22}$, and $\alpha_{12}=\alpha_{21}$.
The explicit forms of the second-order Sonine solutions are too large to be displayed here and will be omitted for the sake of brevity. In particular, since ${\bf j}_{1}^{(1)}=-{\bf j}_{2}^{(1)}$, $D^*[2]$ must be symmetric while $D_{p}^*[2]$ and $D^{\prime \ast}[2]$ must be antisymmetric with respect to the exchange $1\leftrightarrow 2$. We have checked that our expressions verify the above symmetry properties.

It must be noted again that \emph{all} the above expressions have the power to be explicit; that is they are explicitly given in terms of the parameters of the mixture.\cite{footnote}  Since our theory does not involve numerical solutions the diffusion transport coefficients can be evaluated within very short computing times. \cite{footnote}

It is quite apparent that the influence of the parameters of the mixture on the second Sonine approximations is rather complicated, given the large number of parameters involved in the system. Thus, in order to show more clearly the dependence on each parameter on diffusion, it is instructive to consider first some simple cases.

\subsection{Some special limits}

Let us first consider the first Sonine approximations $D^*[1]$, $D_p^*[1]$, and $D^{'*}[1]$. They can be obtained from the general results by taking $a_{i,2}=b_{i,2}=c_{i,2}=0$. In this case, one gets
\begin{equation}
D^{\ast}[1]=\left(\nu ^{\ast}-\frac{1}{2}\zeta^{\ast
}\right)^{-1}\left[ \left( \frac{\partial }{\partial
x_{1}}x_{1}\gamma_{1}\right) _{p,T}+\left( \frac{\partial \zeta
^{\ast }}{\partial x_{1}} \right) _{p,T}\left( 1-\frac{\zeta
^{\ast}}{2\nu ^{\ast }}\right) D_{p}^{\ast }[1]\right],
\label{3.25}
\end{equation}
\begin{equation}
D_{p}^{\ast}[1]=x_{1}\left( \gamma _{1}-\frac{\mu}{x_2+\mu x_1}
\right) \left( \nu ^{\ast }-\frac{3}{2}\zeta ^{\ast }+\frac{\zeta
^{\ast 2}}{ 2\nu ^{\ast }}\right) ^{-1}, \label{3.26}
\end{equation}
\begin{equation}
D^{\prime\ast}[1]=-\frac{\zeta ^{\ast }}{2\nu ^{\ast }}D_{p}^{\ast}[1], \label{3.27}
\end{equation}
where $\zeta^*\equiv \zeta^{(0)}/\nu_0$ and $\nu^*$ is given by Eq.\ \eqref{b1}.
The temperature ratio $\gamma\equiv T_1/T_2$ is determined from the condition $\zeta_1^*=\zeta_2^{*}=\zeta^{*}$, where the partial cooling rates $\zeta_i^{*}$ are given by Eq.\ \eqref{b13}. The expressions \eqref{3.25}--\eqref{3.27} agree with those derived in previous works. \cite{GD02,GM07}

Another interesting situation is the case of mechanically equivalent particles ($m_1=m_2$, $\sigma_1=\sigma_2$, $\alpha_{11}=\alpha_{22}=\alpha_{12}\equiv \alpha$). In this simple situation, as expected, our results yield $D_p^*[2]=D^{'*}[2]=0$ and
\begin{equation}
\label{3.30}
D^*[2]=D^*[1]\frac{1+\alpha}{d}
\frac{12\alpha^2+3(2d-3)\alpha+8+10d}{12\alpha^3+(6d-5)\alpha^2+(16d+1)\alpha+10d+12},
\end{equation}
where the first Sonine approximation $D^*[1]$ is simply
\begin{equation}
\label{3.31}
D^*[1]=\frac{2\Gamma\left(\frac{d}{2}\right)}{\pi^{\frac{d}{2}-1}}\frac{d}{(1+\alpha)^2}.
\end{equation}
As expected, the expression of the self-diffusion coefficient $D^*[2]$ holds for any relative number of tagged particles since it is independent of $x_1$. Equation \eqref{3.31} coincides with previous results for the self-diffusion coefficient. \cite{BRCG00}

Let us consider finally the tracer limit, namely, we assume that the concentration of one of the species (say for instance, species 1) is negligible ($x_1\to 0$). In this limit, a careful analysis of the matrix equation \eqref{a28} defining the Sonine coefficients $a_{ij}$, $b_{ij}$, and $c_{ij}$ shows that $a_{22}=0$ and the coefficients $a_{11}$ (which defines the diffusion coefficient $D$ through Eq.\ \eqref{3.19}) and $a_{12}$ are decoupled from the remaining 6 Sonine coefficients. Moreover, the coefficients $b_{22}$ and $c_{22}$ associated with the excess component also verify an autonomous set of equations so that, the coefficients $b_{11}$ (which defines the pressure diffusion coefficient $D_p$ through Eq.\ \eqref{3.19}) and $c_{11}$ (which defines the thermal diffusion coefficient $D'$ through Eq.\ \eqref{3.19}) can be given in terms of $b_{22}$ and $c_{22}$. The corresponding expressions for $D^*[2]$, $D_p^*[2]$, and $D^{'*}[2]$ coincide with those obtained previously \cite{GF09} by following an independent route. In particular, the explicit expression of the tracer diffusion coefficient $D^*[2]$ is
\begin{equation}
\label{3.32}
D^*[2]=\gamma \frac{\nu_{11}^*-\frac{3}{2}\zeta^*}{(\nu^*-\frac{1}{2}\zeta^*)(\nu_{11}^*-\frac{3}{2}\zeta^*)-
\omega_{12}^*\tau_{11}^*},
\end{equation}
where $\nu^*$ is given by Eq.\ \eqref{b1} with $x_1=0$ and the quantities $\nu_{11}^*$, $\omega_{12}^*$ and $\tau_{11}^*$ are defined in the Appendices \ref{appA} and \ref{appB}.

All the above limits confirm the self-consistency of the results derived in this paper for the second Sonine approximation to the diffusion coefficients $D$, $D_p$, and $D'$ of a granular binary mixture.

\section{Comparison with DSMC results}
\label{sec4}

Needless to say, the improvement of the analytical results by considering the second Sonine approximation for the diffusion coefficients is not completely guaranteed unless the Sonine polynomial expansion is convergent.
The analysis of higher-order Sonine corrections to the transport coefficients for granular gases and the convergence the Sonine polynomial expansion is a very difficult mathematical problem.
Thus, the works devoted to this issue in granular systems are more scarce than for ordinary gases.
For instance, the analysis of the transport properties for dense binary mixtures have been studied  and it was observed that with one tracer component
($x_1\to 0$) the convergence of the Sonine expansion improves with increasing values of
the mass ratio $m_1/m_2$. \cite{MC84} In this Section, we will compare the first and second Sonine approximations
of the mutual diffusion coefficient $D$ with computer simulation results obtained by numerically solving
the Boltzmann equation by means of the DSMC method. \cite{GF09} As in previous studies, \cite{GM04,GM07,GF09} due
to the difficulties for measuring the coefficient $D$ for general values of the mass ratio and the mole
fraction, we will consider the self-diffusion ($m_1=m_2$) and tracer diffusion ($x_1\to 0$) coefficients.
However, in order to cover more general systems than those considered in our previous simulations, \cite{GM04,GM07,GF09}
we will assume that $\alpha_{12}\neq \alpha_{22}$ when the intruder and the gas particles are mechanically
different.
\begin{figure*}
\begin{center}
\begin{tabular}{lr}
\resizebox{8cm}{!}{\includegraphics{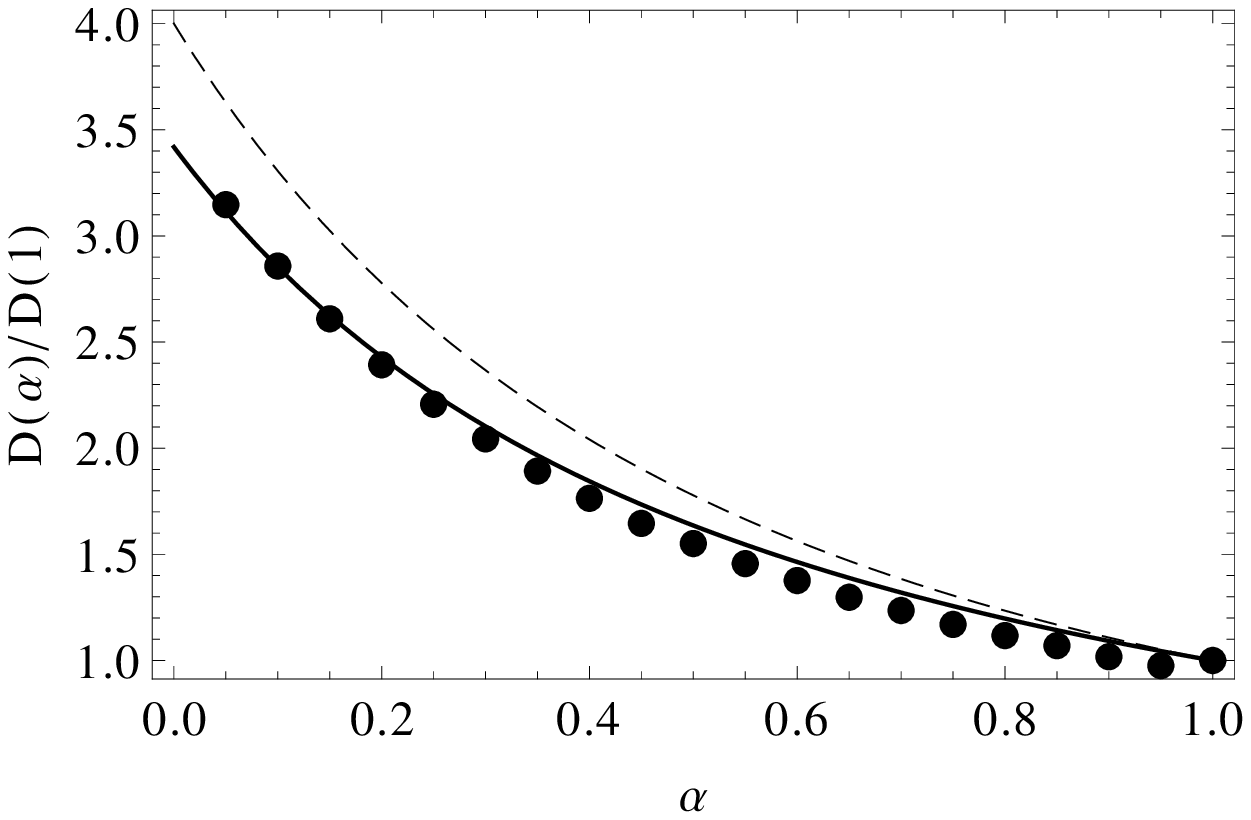}}&\resizebox{8cm}{!}
{\includegraphics{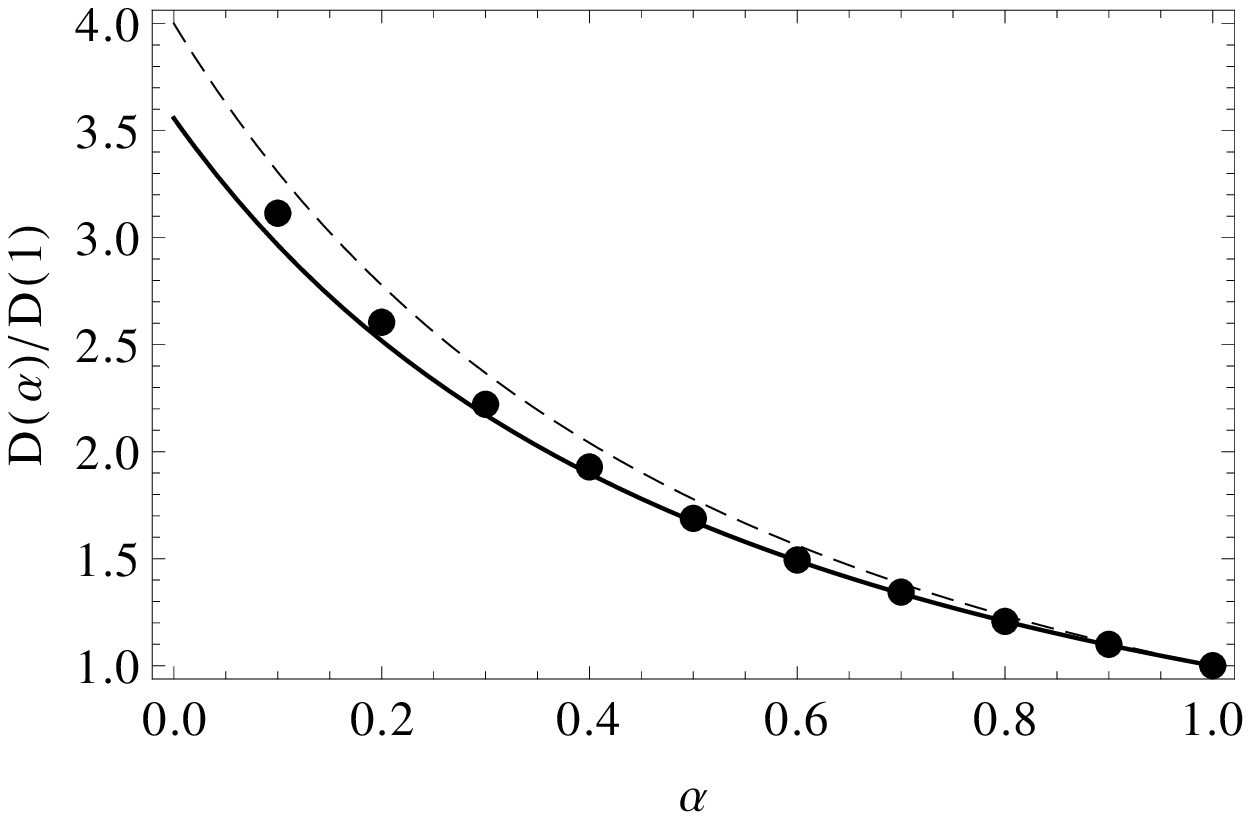}}
\end{tabular}
\end{center}
\caption{Plot of the (reduced) self-diffusion coefficient $D(\alpha)/D(1)$ as a function of the coefficient of restitution $\alpha$ as given by the first Sonine approximation (dashed line), the second Sonine approximation (solid line), and Monte Carlo simulations (symbols). Here, $D(1)$ is the elastic value of the self-diffusion coefficient consistently obtained in each approximation. The left panel is for hard disks ($d=2$) while the right panel is for hard spheres ($d=3$).
\label{fig1sim}}
\end{figure*}

The adaption of DSMC method to analyze binary granular mixtures has been described in previous works (see, for instance, Refs.\ \onlinecite{GM04} and \onlinecite{MG02}), so that here we shall only mention some aspects related to the diffusion of impurities in a granular gas under HCS. In the tracer limit ($n_1\ll n_2$), during our simulations collisions 1-1 are not considered, and when a collision 1-2 takes place, the post-collisional velocity obtained from the scattering rule is only assigned to the tracer particle (species 1). According to this scheme, the numbers of particles have simply a statistical meaning and can be arbitrarily chosen.

The DSMC method for our problem has two steps that are repeated in each time iteration. \cite{GF09} In the first step, the system (tracer and gas particles) evolves from the initial state to the HCS. In the second step, the system is assumed to be in the HCS and then the diffusion coefficient $D(t)$ is measured from the mean square displacement of the impurity as
\begin{equation}
\label{4.1}
D(t)=\frac{n_2}{2d \delta t}\left[\langle |{\bf r}(t+\delta t)-{\bf r}(0)|^2 \rangle-
\langle |{\bf r}(t)-{\bf r}(0)|^2 \rangle\right].
\end{equation}
Here, $|{\bf r}(t)-{\bf r}(0)|$ is the distance traveled by the impurity from $t=0$ until time $t$, $t=0$ being the beginning of the second step. Moreover, $\langle \cdots \rangle$ denotes the average over the $N$ impurities and $\delta t$ is the time step. In our simulations, we have typically taken a time step $\delta t=2.5\times10^{-4}\nu^{-1}$ and $N=2\times 10^6$ simulated particles for each species. Here, $\nu=n_2 \sigma^{d-1}\sqrt{2T/m_2}$ is an effective collision frequency for gas particles.

\begin{figure}[htbp]
\begin{center}
\resizebox{8cm}{!}{\includegraphics{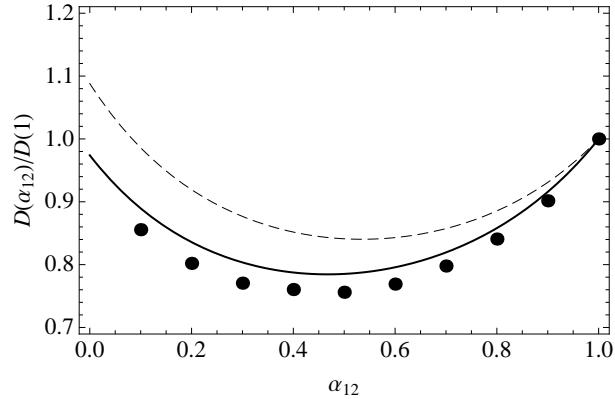}}
\end{center}
\caption{Plot of the (reduced) mutual diffusion coefficient $D(\alpha_{12})/D(1)$ versus the coefficient of restitution $\alpha_{12}$ in the tracer limit ($x_1\to 0$) for a granular gas of hard spheres with $\omega=1/2$, $\mu=1/4$ and $\alpha_{22}=0.5$. The dashed and solid lines are first and second Sonine approximations, respectively, while the symbols are the Monte Carlo simulation results. Here, $D(1)$ is the elastic value of the mutual diffusion coefficient consistently obtained in each approximation. \label{fig3sim}}
\end{figure}
\begin{figure}[htbp]
\begin{center}
\resizebox{8cm}{!}{\includegraphics{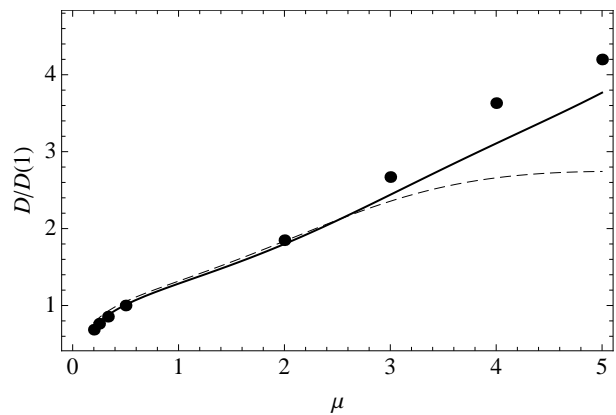}}
\end{center}
\caption{Plot of the (reduced) mutual diffusion coefficient $D/D(1)$ as a function of the mass ratio $\mu$ in the tracer limit ($x_1\to 0$) for a granular gas of hard spheres with $\omega=1/2$ and a (common) coefficient of restitution  $\alpha\equiv\alpha_{22}=\alpha_{12}=0.5$. The dashed and solid lines are first and second Sonine approximations, respectively, while the symbols are the Monte Carlo simulation results. Here, $D(1)$ is the elastic value of the mutual diffusion coefficient consistently obtained in each approximation.} \label{fig4sim}
\end{figure}

We will consider first the self-diffusion coefficient, which is independent of the mole fraction $x_1$ [see Eqs.\ \eqref{3.30} and \eqref{3.31}]. The simulation data obtained from DSMC method along with both Sonine approximations for the reduced coefficient $D(\alpha)/D(1)$ are presented in Fig.\ \ref{fig1sim} for disks ($d=2$) and spheres ($d=3$). Here, $D(1)$ refers to the elastic value of the self-diffusion coefficient consistently obtained in each Sonine approximation. The data corresponding to $d=3$ for $\alpha \geq 0.5$ and $d=2$ for $\alpha \geq 0.6$ were reported in Refs.\ \onlinecite{GM04} and \onlinecite{GM07}, respectively, while those corresponding to $d=3$ and $d=2$ for $\alpha \leq 0.5$ have been obtained in this work. It is quite apparent that the first Sonine approximation performs well for not strong values of dissipation, but the agreement between theory and simulation improves over the complete range of values of the coefficient of restitution when the second Sonine approximation is considered (especially for hard disks). This confirms again the accuracy of the second Sonine approach even for quite extreme values of dissipation.

Consider now the situation in which impurities and particles of the gas are mechanically different (i.e., they can differ in size, mass and coefficients of restitution). Although not shown here, as expected, \cite{GM04} comparison between theory and simulation shows that the Sonine polynomial expansion exhibits a better convergence (namely, although both Sonine approximations compare well with numerical results, the second is better) when the impurity is heavier and/or larger than the gas particles while this convergence is worsen as $\mu$ and/or $\omega$ significantly decreases. These findings agree with the conclusions obtained for elastic collisions. \cite{MC84} To illustrate this behavior, Fig.\ \ref{fig3sim} shows the dependence of the ratio $D(\alpha_{12})/D(1)$ on the coefficient of restitution $\alpha_{12}$ for hard spheres with $\omega=1/2$, $\mu=1/4$ and $\alpha_{22}=0.5$. The present comparison complements previous results \cite{GM04,GM07,GF09} reported for the special case $\alpha_{12}=\alpha_{22}$. We observe that the first Sonine approximation clearly overestimates the simulation results while the second Sonine approximation to $D(\alpha_{12})$ exhibits good agreement. On the other hand, the quantitative discrepancies between the second Sonine solution and simulation data are larger than those observed for the self-diffusion problem (see Fig.\ \ref{fig1sim}), especially for strong dissipation. Thus, one perhaps would have to consider the third Sonine correction to obtain a better prediction for the diffusion coefficient.

We explore now the influence of the mass ratio $\mu$ on the accuracy of the two first Sonine approximations. Figure \ref{fig4sim} shows the ratio $D/D(1)$ versus the mass ratio $\mu$ for hard spheres with $\omega=1/2$ and a (common) coefficient of restitution  $\alpha\equiv\alpha_{22}=\alpha_{12}=0.5$. We find that the second Sonine approximation $D[2]$ differs form the first Sonine approximation $D[1]$ as the mass ratio $\mu$ is varied. For the system studied in Fig.\ \ref{fig4sim}, the disagreement between both approaches turns out to be significant when the impurity is heavier than the gas particles. Thus, for instance when $\mu=5$, the first Sonine approximation to the ratio $D/D(1)$ differs by 26\% from the second Sonine approximation. The comparison with simulation data shows again that the theoretical predictions are clearly improved when one takes the second Sonine solution (up to $20\%$ of improvement compared to the first Sonine approximation). However, the quantitative differences between the second Sonine solution and DSMC results seem to increase as the mass ratio increases. In this case, as in Fig.\ \ref{fig3sim}, one should consider higher-order terms in the Sonine polynomial expansion to get a more accurate approach. We want also to remark that we have also considered other systems (see for instance, Figs.\ 8 and 9 of Ref.\ \onlinecite{GM04} and Figs.\ 4 and 5 of Ref.\ \onlinecite{GF09}) where the improvement of the second Sonine approximation to $D$ over the first Sonine approximation is much more significant than the one observed in Figs.\ \ref{fig1sim}, \ref{fig3sim}, and \ref{fig4sim}.

The results reported in this Section confirm again the reliability of the second Sonine approximation for the mutual diffusion coefficient $D$, at least in the cases of self-diffusion and tracer limit. Unfortunately, the lack of available simulation data for finite mole fraction prevent us to assess the reliability of the second Sonine solution to $D$ beyond the tracer limit. The fact that the second Sonine expression for $D$ in the self-diffusion problem (which holds for any value of $x_1$) compares quite well with DSMC results suggests that the good agreement found for $x_1\to 0$ would be also kept for arbitrary values of the mole fraction, even when both species are mechanically different. More simulations are needed to support the above expectation.
\begin{figure}[htbp]
\begin{center}
\resizebox{9cm}{!}{\includegraphics{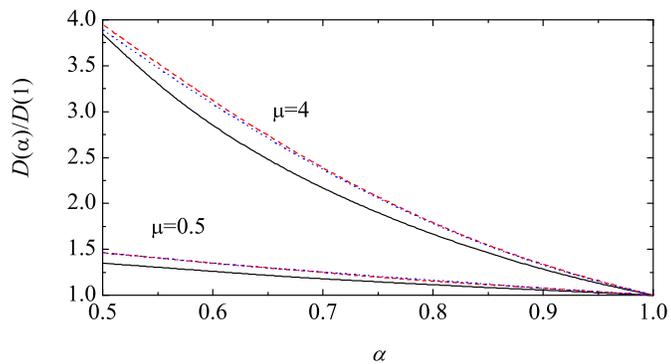}}
\end{center}
\caption{(color online) Plot of the reduced coefficient $D(\alpha)/D(1)$ as a function of the (common) coefficient of restitution $\alpha$ for hard spheres with $x_1=0.2$, $\sigma_1=\sigma_2$ and two different values of the mass ratio $\mu\equiv m_1/m_2$. The solid lines correspond to the results obtained from the second Sonine approximation, the dashed lines refer to the (standard) first Sonine approximation and the dotted lines correspond to the modified first Sonine approximation. Here, $D(1)$ is the elastic value of $D$ consistently obtained in each approximation. \label{figdif}}
\end{figure}
\begin{figure}[htbp]
\begin{center}
\resizebox{9cm}{!}{\includegraphics{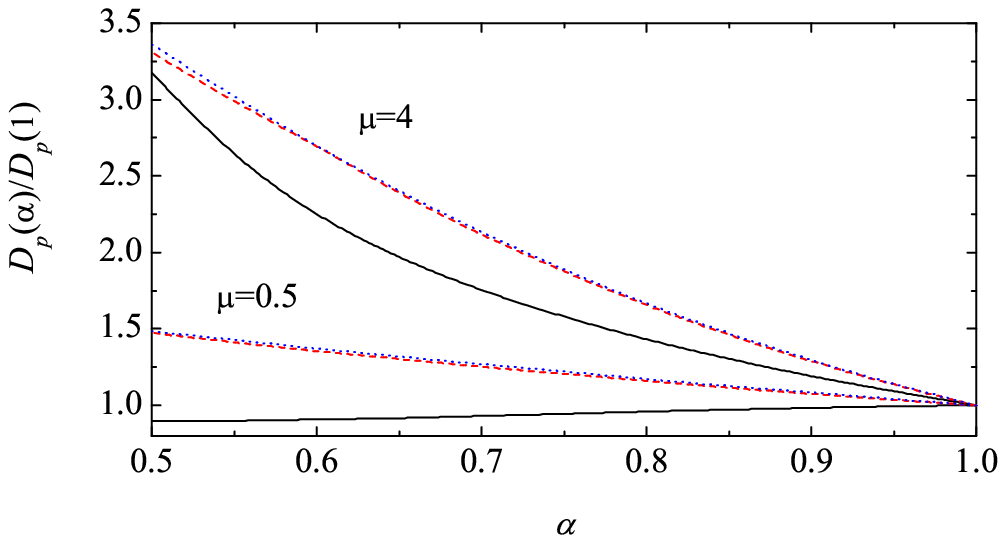}}
\end{center}
\caption{(color online) Plot of the reduced coefficient $D_p(\alpha)/D_p(1)$ as a function of the (common) coefficient of restitution $\alpha$ for hard spheres with $x_1=0.2$, $\sigma_1=\sigma_2$ and two different values of the mass ratio $\mu\equiv m_1/m_2$. The solid lines correspond to the results obtained from the second Sonine approximation, the dashed lines refer to the (standard) first Sonine approximation and the dotted lines correspond to the modified first Sonine approximation. Here, $D_p(1)$ is the elastic value of $D_p$ consistently obtained in each approximation. \label{figdp}}
\end{figure}
\begin{figure}[htbp]
\begin{center}
\resizebox{9cm}{!}{\includegraphics{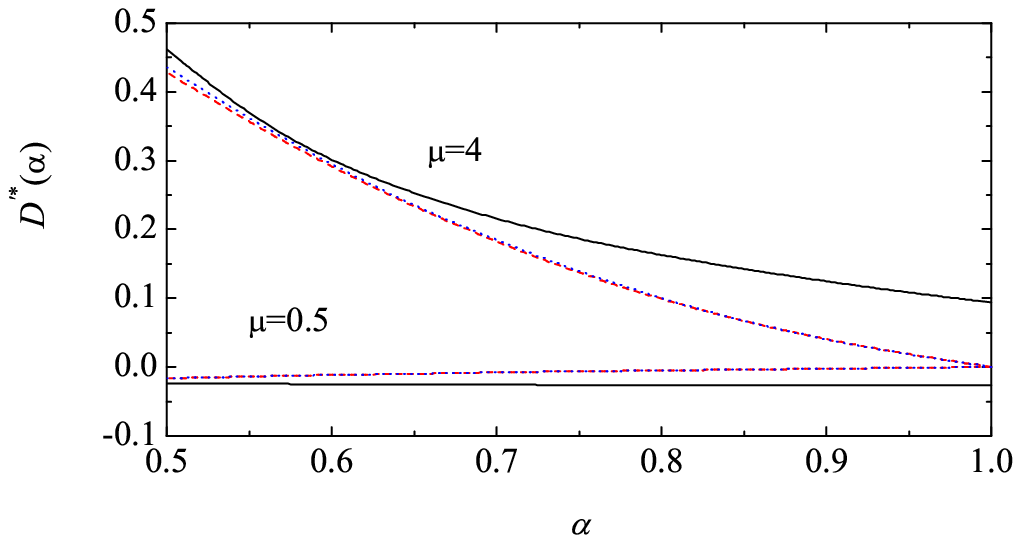}}
\end{center}
\caption{(color online) Plot of the reduced coefficient $D^{'*}(\alpha)$ as a function of the (common) coefficient of restitution $\alpha$ for hard spheres with $x_1=0.2$, $\sigma_1=\sigma_2$ and two different values of the mass ratio $\mu\equiv m_1/m_2$. The solid lines correspond to the results obtained from the second Sonine approximation, the dashed lines refer to the (standard) first Sonine approximation and the dotted lines correspond to the modified first Sonine approximation.\label{figdprime}}
\end{figure}
\begin{figure}[htbp]
\begin{center}
\resizebox{9cm}{!}{\includegraphics{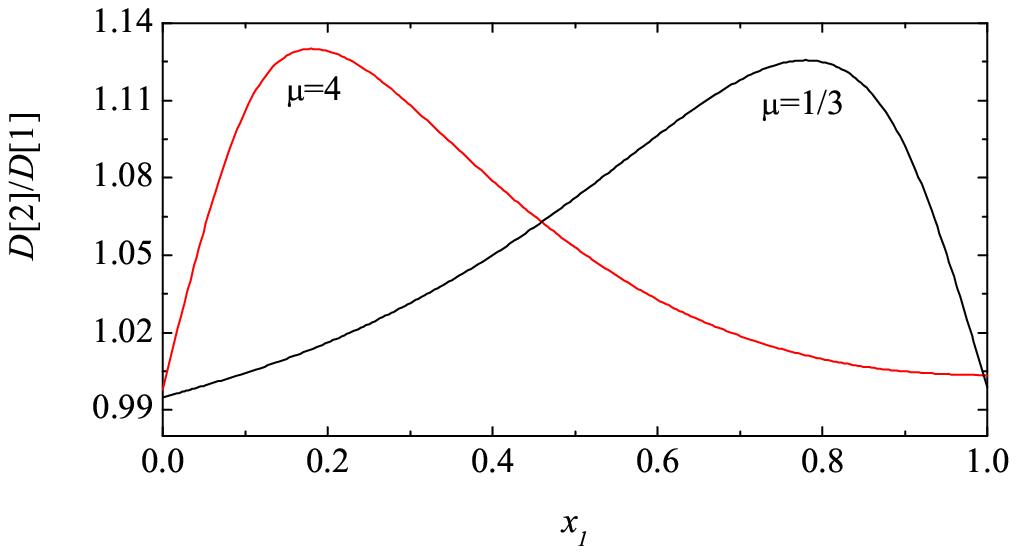}}
\end{center}
\caption{(color online)The ratio of the second and first Sonine approximations $D[2]/D[1]$ to the mutual
diffusion coefficient versus the mole fraction $x_1$ for $\omega=1$, $\alpha=0.8$ and two values of the mass ratio ($\mu=4$ and $\mu=1/3$).} \label{figdifx1}
\end{figure}
\begin{figure}[htbp]
\begin{center}
\resizebox{9cm}{!}{\includegraphics{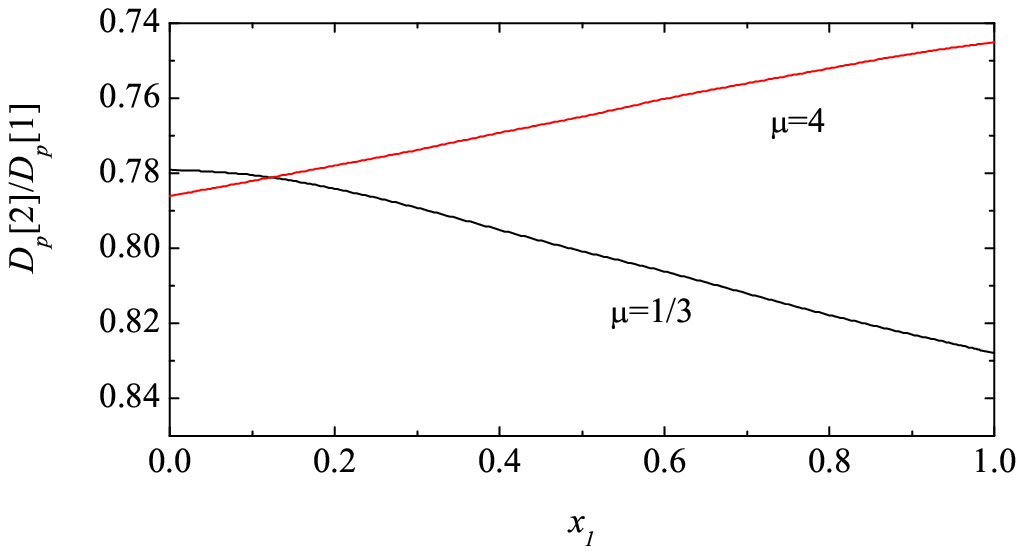}}
\end{center}
\caption{(color online)The ratio of the second and first Sonine approximations $D_p[2]/D_p[1]$ to the pressure
diffusion coefficient versus the mole fraction $x_1$ for $\omega=1$, $\alpha=0.8$ and two values of the mass ratio ($\mu=4$ and $\mu=1/3$).} \label{figdpx1}
\end{figure}
\begin{figure}[htbp]
\begin{center}
\resizebox{9cm}{!}{\includegraphics{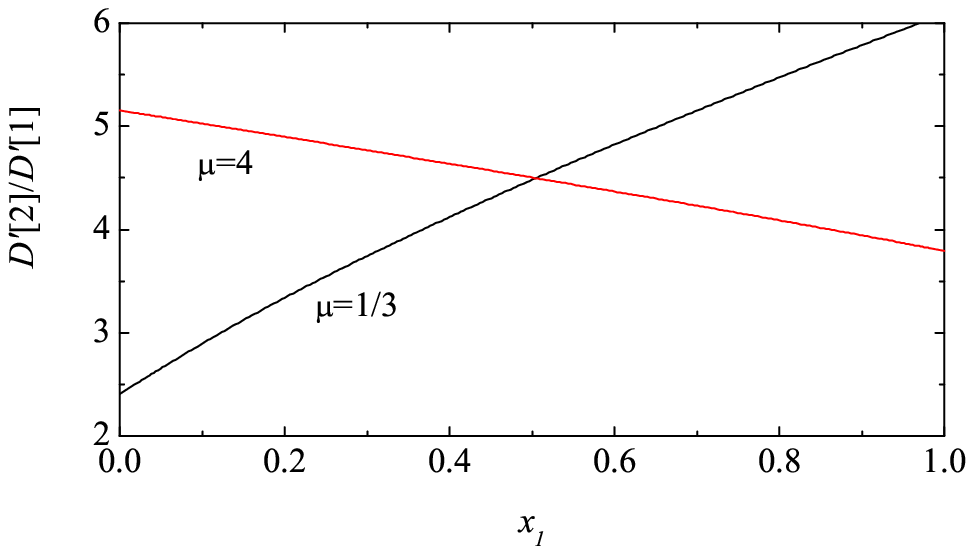}}
\end{center}
\caption{(color online)The ratio of the second and first Sonine approximations $D'[2]/D'[1]$ to the thermal
diffusion coefficient versus the mole fraction $x_1$ for $\omega=1$, $\alpha=0.8$ and two values of the mass ratio ($\mu=4$ and $\mu=1/3$).} \label{figdprimex1}
\end{figure}

\section{Dependence of the diffusion coefficients on the parameters of the mixture}
\label{sec5}

Once the reliability of the second Sonine solution to the mutual diffusion coefficient $D$ has been confirmed in the previous Section, our goal now is to provide a systematic study of the dependence of the complete set of diffusion coefficients $D$, $D_p$ and $D'$ on the parameter space of the system. However, the first and second Sonine approximations to the (reduced) transport coefficients of the granular binary mixture depend on many parameters: $\left\{x_1, m_1/m_2, \sigma_1/\sigma_2, \alpha_{11}, \alpha_{22}, \alpha_{12}\right\}$.  Also, to reduce the number of independent parameters, the simplest case of a \emph{common} coefficient of restitution ($\alpha_{11}=\alpha_{22}=\alpha_{12}\equiv \alpha$) and a \emph{common} diameter ($\sigma_1=\sigma_2$) is considered. The latter assumption is justified because the dependence of $D^*$, $D_p^*$ and $D^{'* }$ on the diameter ratio $\omega$ is very weak. Moreover, henceforth we only analyze the physical case of hard spheres ($d=3$) and so, the parameter space is reduced to three quantities: $\left\{x_1, m_1/m_2, \alpha \right\}$.

The first and second Sonine approximations of the (reduced) transport coefficients $D(\alpha)/D(1)$, $D_p(\alpha)/D_p(1)$, and $D^{'*}(\alpha)$ are plotted in Figs.\  \ref{figdif}, \ref{figdp}, and \ref{figdprime}, respectively,  for $x_1=0.2$ and two values of the mass ratio $\mu$. The diffusion coefficients have been reduced with respect to their elastic values (consistently obtained in each Sonine approximation), except the thermal diffusion coefficient $D'$ since it vanishes for elastic collisions when one considers the first Sonine approximation. In this latter case, we have plotted the reduced coefficient $D^{'*}$ defined by the third relation in Eq.\ \eqref{3.23}. For the sake of comparison, we have also included the results derived from a modified version of the first Sonine approximation. \cite{GFM09}
This approach consists of replacing the Maxwellian distribution in the first Sonine solution by the HCS distribution. Figure \ref{figdif} shows the $\alpha$-dependence of the mutual diffusion coefficient obtained from the three different approximations (standard and modified first Sonine approximation and the second Sonine approximation) for two mass ratios. We observe that the first Sonine approximations capture relatively well the effect of dissipation on the mutual diffusion coefficient since the three approaches show a monotonic increase of $D$ with decreasing $\alpha$ in all cases. On the other hand, at a more quantitative level, both first Sonine solutions overestimate slightly the predictions of the second Sonine approach. In any case, the convergence of the Sonine expansion for this transport coefficient seems to be quite good, at least for not quite extreme values of mass and/or diameter ratios.

We consider now the pressure diffusion coefficient $D_p(\alpha)$. This is plotted in Fig.\ \ref{figdp} for the same cases as in Fig.\ \ref{figdif}. In contrast to the case of the mutual diffusion coefficient, when the defect species is lighter than the excess component, the dependence of $D_p$ on the coefficient of restitution predicted by the first Sonine approximation ($D_p$ increases with decreasing $\alpha$) differs from the one obtained from the more refined second Sonine solution ($D_p$ decreases with decreasing $\alpha$). At a quantitative level, the first Sonine approximations overestimate again the second Sonine results for both values of the mass ratio, being the differences between both Sonine solutions more pronounced when $\mu<1$. In fact, at $\alpha=0.5$, the discrepancies between the first and second Sonine approximations are about 4 \% for $\mu=4$ while they are about 63 \% for $\mu=0.5$. The dependence of the thermal diffusion coefficient $D^{'*}$ on $\alpha$ is shown in Fig.\ \ref{figdprime}. Note that, in the elastic limit, the first Sonine approximation to $D^{'*}$ vanishes while the second Sonine approximation is in general different from zero. We observe that both Sonine results tend to approach each other as the dissipation increases. In particular, the dependence of $D^{'*}$ on the coefficient of restitution predicted by the the first and second Sonine approximations is very weak  when $\mu<1$ (in fact it is practically zero) while the coefficient increases clearly with dissipation in the opposite case ($\mu>1$). In comparison with the results obtained for $D_p$, the convergence of the Sonine solution for $D'$ is better than that of the pressure diffusion coefficient, specially for strong dissipation. It must be noticed that the differences between the standard and modified first Sonine approximations \cite{GFM09} are quite small in the region of collisional dissipation considered. Although not shown here, similar conclusions can be drawn when one considers other values for the mass and size ratios.

As said in the Introduction, the results derived in this paper extend previous studies (on both Sonine approximations) on the diffusion coefficients in the tracer limit ($x_1\to 0$). \cite{GM04,GF09} Thus, one of the goals here is to assess the effect of finite concentration on the ratios of the second and first Sonine approximations to the diffusion transport coefficients. Figures \ref{figdifx1}, \ref{figdpx1}, and \ref{figdprimex1}  shows the ratios $D[2]/D[1]$, $D_p[2]/D_p[1]$ and $D'[2]/D'[1]$, respectively, versus the concentration $x_1$ for $\omega=1$, $\alpha=0.8$ and two (disparate) values of the mass ratio $\mu$. The impact of composition on the above ratios is in general significant. While the ratio $D[2]/D[1]$ has a non monotonic dependence of $x_1$, the corresponding ratios for the pressure and thermal diffusion coefficients exhibit a monotonic dependence with $x_1$. The second Sonine approximation to the diffusion coefficients differs clearly from its first Sonine approximation, specially in the case of the thermal diffusion coefficient (we observe for instance up to a $500\%$ difference for $D'$ in Fig.\ \ref{figdprimex1}).

\section{Thermal diffusion segregation}
\label{sec6}

As an application of the previous results, this section is devoted to the study of segregation driven by a thermal gradient in granular binary mixtures. This is one of the most interesting problems appearing in multicomponent mixtures and it has been widely analyzed in the past  for ordinary gases and liquids. \cite{Lambda} On the other hand, much less is known in the case of granular mixtures, although some progress has been made in the past few years in the tracer limit case ($x_1\to 0$). \cite{G06,G08,GF09,G11, BRM05,BKD11} Here, we analyze thermal diffusion for arbitrary concentrations but restricted to the case of \emph{dilute} granular systems.

We consider a granular binary mixture enclosed between two plates at different temperatures. In a non-convecting steady state ($\textbf{u}=\textbf{0}$) with gradients only along the orthogonal direction to the plates ($z$ axis), the amount of segregation parallel to the thermal gradient may be characterized by the thermal diffusion factor $\Lambda$. This quantity measures the separation of components caused by the temperature gradient. The factor $\Lambda$ is defined as \cite{KCM87,AW98}
\begin{equation}
\label{6.1}
-\Lambda\frac{\partial \ln T}{\partial z} =\frac{\partial}{\partial z}\ln
\left(\frac{n_1}{n_2}\right).
\end{equation}
Let us assume henceforth that $\sigma_1\geq \sigma_2$ and that the bottom plate is hotter than the top plate ($\partial_z T<0$). In this case and assuming that $\Lambda$ is constant over the relevant ranges of temperature and composition, when $\Lambda>0$ the larger particles $1$ tend to rise with respect to the smaller particles $2$ (i.e., $\partial_z (n_1/n_2)>0$). In the opposite case, when $\Lambda<0$ the larger particles fall with respect to the smaller particles (i.e., $\partial_z (n_1/n_2)<0$). Although gravity is absent in our description, the former situation ($\Lambda>0$) will be referred here to as the Brazil-nut effect (BNE) while the latter ($\Lambda<0$) will be called as the reverse Brazil-nut effect (RBNE).

In the case of granular mixtures thermal diffusion can appear in vibrated systems even in the absence of an imposed temperature gradient, as a consequence of the inelasticity of collisions. In this case, energy of the grains decays away from the vibrating wall, giving rise to a (granular) temperature gradient.
However, it is known for vertically vibrated granular systems \cite{Brey2,BT02} that after
the decrease in the value of the granular temperature as a function of height above the floor, the
temperature profile possesses a minimum above which the temperature increases as a function of height.
Therefore, given that we have assumed $\partial_z T<0$ in Eq.\ \eqref{6.1}, then our segregation criterion can
be useful for physical situations \cite{VU09,VSG13} where the minimum in the temperature profile
is not achieved or is very close to the top of the sample.

Since no shearing flows are present in the problem, the pressure tensor $P_{ij}=p\delta_{ij}$ and so the momentum balance equation \eqref{2.7} yields simply $\partial_z p=0$. Moreover, according to Eq.\ \eqref{2.6}, $j_{1,z}=0$ in the steady state. In the NS hydrodynamic order, $j_{1,z}$ is given by Eq.\ \eqref{1.1} so that the condition $j_{1,z}=0$ (along with $\partial_z p=0$) leads to the relation
\begin{equation}
\label{6.2}
\frac{\partial \ln x_1}{\partial z}=-\frac{\rho^2}{m_1m_2n_1}\frac{D'}{D}\frac{\partial \ln T}{\partial z}.
\end{equation}
The form of $\Lambda$ can be easily obtained from Eqs.\ \eqref{6.1} and \eqref{6.2} and the result is
\begin{equation}
\label{6.3}
\Lambda=\frac{n\rho^2}{m_1m_2n_1n_2}\frac{D'}{D}=\frac{1}{x_1x_2}\frac{D^{'*}}{D^*},
\end{equation}
where use has been made of the reduced expressions \eqref{3.23} for the mutual and thermal diffusion coefficients, respectively. Since the mutual diffusion coefficient D must be positive, the sign of $\Lambda$ is determined by the
sign of the reduced coefficient $D^{'*}$. Consequently, the condition $\Lambda=0$ (which provides the criterion for the BNE/RBNE transition) implies simply
\begin{equation}
\label{6.4}
D^{'*}=0.
\end{equation}
\begin{figure}[htbp]
\begin{center}
\resizebox{9cm}{!}{\includegraphics{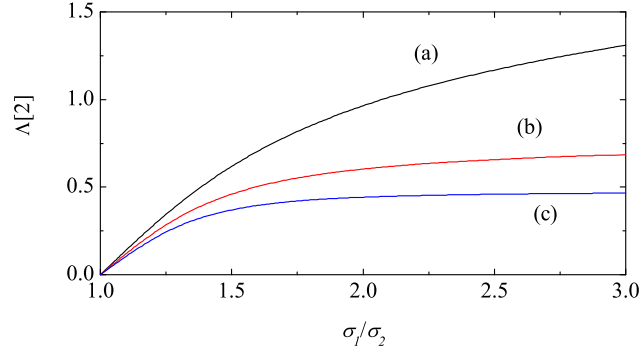}}
\end{center}
\caption{(color online) Plot of the thermal diffusion factor $\Lambda[2]$ obtained from the second Sonine approximation as a function of the diameter ratio $\sigma_1/\sigma_2$ for an ordinary binary mixture ($\alpha_{ij}=1$) of hard spheres when both species have the same mass density ($m_1/m_2=(\sigma_1/\sigma_2)^3$). Three different values of the mole fraction are considered: (a) $x_1=0.2$, (b) $x_1=0.5$, and (c) $x_1=0.8$.} \label{fig3}
\end{figure}
According to Eqs.\ \eqref{3.26} and \eqref{3.27}, the first Sonine approximation to Eq.\ \eqref{6.4} yields the criterion
\begin{equation}
\label{6.5}
\frac{x_1x_2\mu\zeta^*}{(2\nu^{\ast 2}-3\zeta^{\ast}\nu^*+\zeta^{\ast 2})(x_2+x_1\gamma)(x_2+\mu x_1)}
\left(1-\frac{\gamma}{\mu}\right)=0.
\end{equation}
In the elastic limit ($\alpha_{ij}=1$), $\zeta^*=0$ and so, $\Lambda[1]=0$ in the first Sonine approximation. However, away from the dilute gas limit, $\Lambda[1]$ is not zero \cite{KCM87,G11} and segregation appears for ordinary mixtures. In the case of granular mixtures ($\alpha_{ij} \neq 1$), the solution to Eq.\ \eqref{6.5} is simply \cite{G06}
\begin{equation}
\label{6.6}
\frac{m_1}{m_2}=\frac{T_1}{T_2}.
\end{equation}
Note that if one assumes energy equipartition ($T_1=T_2$), then segregation is only predicted for particles that differ in mass, no matter what their diameters may be. It must be emphasized that the criterion \eqref{6.6} compares well with molecular dynamics simulations \cite{BRM05} carried out in the tracer limit ($x_1 \to 0$).

The second Sonine approximation to Eq.\ \eqref{6.4} leads to a much more intricate criterion than Eq.\ \eqref{6.6}. In particular, the results show that $\Lambda[2]\neq 0$ even for elastic collisions ($\alpha_{ij}=1$). This is consistent with the results obtained years ago in Ref.\ \onlinecite{KCM87}. To illustrate this feature, Fig. \ \ref{fig3} shows $\Lambda[2]$ versus $\sigma_1/\sigma_2$ for a binary mixture of hard spheres ($d=3$) constituted by particles of the same mass density ($m_1/m_2=(\sigma_1/\sigma_2)^3$). In this case, $\Lambda [2]$ is always positive and so, the larger particles tend to move towards the cold plate (BNE). It must be remarked again that the second Sonine approximation does predict segregation in the elastic limit whereas the first Sonine approximation does not. Thus, it is expected that the second Sonine solution describes a much better behavior than the first one in the range of small inelasticities.
\begin{figure}[htbp]
\begin{center}
\resizebox{9cm}{!}{\includegraphics{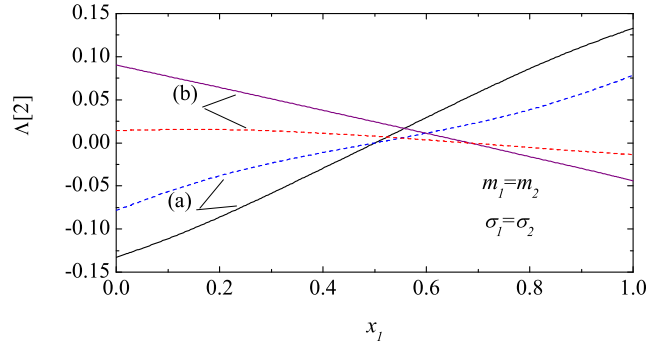}}
\end{center}
\caption{(color online) Plot of the thermal diffusion factors $\Lambda[2]$ and $\Lambda[1]$ as a function of the mole fraction $x_1$ for $m_1=m_2$, $\sigma_1=\sigma_2$ and different values of the coefficients of restitution: (a) $\alpha_{11}=\alpha_{22}=0.5$, $\alpha_{12}=0.9$, and (b) $\alpha_{11}=0.8$, $\alpha_{22}=0.9$, $\alpha_{12}=0.7$. The solid lines correspond to the second Sonine approximation $\Lambda[2]$ while the dashed lines refer to the first Sonine approximation $\Lambda[1]$.} \label{fig4}
\end{figure}
\begin{figure}[htbp]
\begin{center}
\resizebox{9cm}{!}{\includegraphics{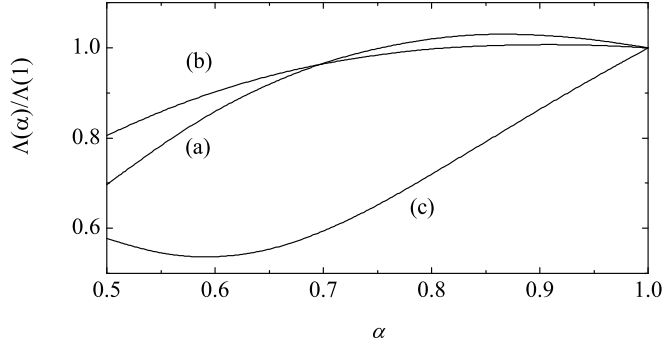}}
\end{center}
\caption{Plot of the second Sonine approximation to the ratio $\Lambda(\alpha)/\Lambda(1)$ as a function of the (common) coefficient of restitution $\alpha$ for $x_1=0.5$, $\sigma_1/\sigma_2=2$ and three different values of the mass ratio: (a)  $m_1/m_2=4$, (b) $m_1/m_2=8$, and (c) $m_1/m_2=1/4$. Here, $\Lambda(1)$ refers to the elastic value of the thermal diffusion factor. } \label{vicente1}
\end{figure}
\begin{figure}[htbp]
\begin{center}
\resizebox{9cm}{!}{\includegraphics{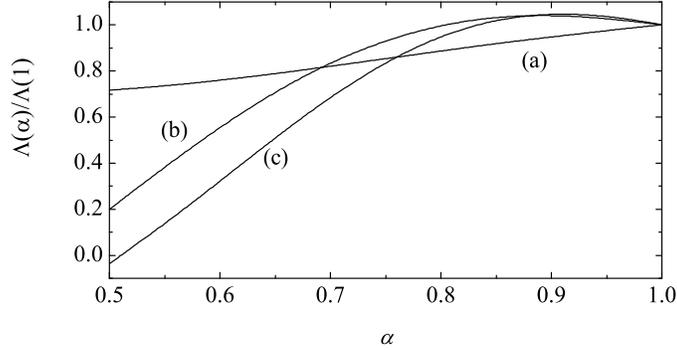}}
\end{center}
\caption{Plot of the second Sonine approximation to the ratio $\Lambda(\alpha)/\Lambda(1)$ as a function of the (common) coefficient of restitution $\alpha$ for $x_1=0.5$, $m_1/m_2=2$ and three different values of the size ratio: (a)  $\sigma_1/\sigma_2=1$, (b) $\sigma_1/\sigma_2=3$, and (c) $\sigma_1/\sigma_2=5$. Here, $\Lambda(1)$ refers to the elastic value of the thermal diffusion factor.} \label{vicente2}
\end{figure}

Another interesting limit case corresponds to the situation in which segregation is only induced by inelasticity, namely, when one considers a binary mixture whose constituents differ \emph{only} by their respective coefficients of restitution. This situation has been theoretically studied \cite{SGNT06} from the Boltzmann equation and it has been also confirmed \cite{BEGS08} by molecular dynamics simulations of two-dimensional binary mixtures. In order to analyze this effect, Fig.\ \ref{fig4} shows the thermal diffusion factor $\Lambda$ versus the mole fraction $x_1$ when $m_1=m_2$, $\sigma_1=\sigma_2$ and different values of the coefficients of restitution. As expected, \cite{SGNT06} we observe that segregation can occur due to inelasticity alone. Notice also that for the cases represented in Fig. 11, the first and second Sonine approximation have differences of about $800\%$ (note the tracer limit $x_1\to 0$ for the $\alpha_{11}=0.8$ curves. Also, in both systems there is a change in the sign of $\Lambda$ at a given critical value $x_{1,c}$ of the composition $x_1$. Although the form of $\Lambda$ differs in the first and second Sonine approximations, the value $x_{1,c}$ for each mixture is (practically) the same in both Sonine predictions. In the case (a), $x_{1,c}=0.5$ due to symmetry considerations.

Apart from the above limit situations, the dependence of $\Lambda$ on the parameter space is quite intricate. To assess the effect of inelasticity in collisions on thermal diffusion factor, we normalize $\Lambda(\alpha)$ with respect to its value in the elastic limit $\Lambda(1)$. Moreover, we consider again the physical case of hard spheres ($d=3$) with a common coefficient of restitution ($\alpha_{ij}=\alpha$) and only the second Sonine approximation to $\Lambda$ will be plotted. Figure \ref{vicente1} shows $\Lambda(\alpha)/\Lambda(1)$ as a function of $\alpha$ for an equimolar mixture ($x_1=0.5$) with  $\sigma_1/\sigma_2=2$ and three different values of the mass ratio $m_1/m_2$. We observe that the impact of collisional dissipation on thermal diffusion is in general quite significant. It is apparent that thermal diffusion is partly concealed by inelasticity since $|\Lambda(\alpha)|<|\Lambda(1)|$. In addition, Fig.\ \ref{vicente1} also shows that the dependence of $\Lambda$ on the mass ratio is non monotonic when the mass ratio is larger than one: while the magnitude of the ratio $\Lambda(\alpha)/\Lambda(1)$ decreases with increasing the mass ratio at moderate inelasticity (say for instance, $\alpha \gtrsim 0.9$), the opposite happens at smaller values of the coefficient of restitution. The $\alpha$-dependence of the ratio $\Lambda(\alpha)/\Lambda(1)$ is also plotted in Fig.\ \ref{vicente2} for $x_1=0.5$, $m_1/m_2=2$ and three different values of the diameter ratio $\sigma_1/\sigma_2$. As happens in Fig.\ \ref{vicente1}, the influence of dissipation on thermal diffusion is again quite significant, especially when the sizes of both species are very disparate. In addition, in the case $\sigma_1/\sigma_2=5$, we also observe that there is a change of the sign of $\Lambda$ for high inelasticity. Thus, for this system, while the larger particles tend to accumulate at the top of the sample when both species collide elastically, the opposite happens for high dissipation and the larger particles fall with respect to the smaller ones.
\begin{figure*}
\begin{center}
\begin{tabular}{lr}
\resizebox{8cm}{!}{\includegraphics{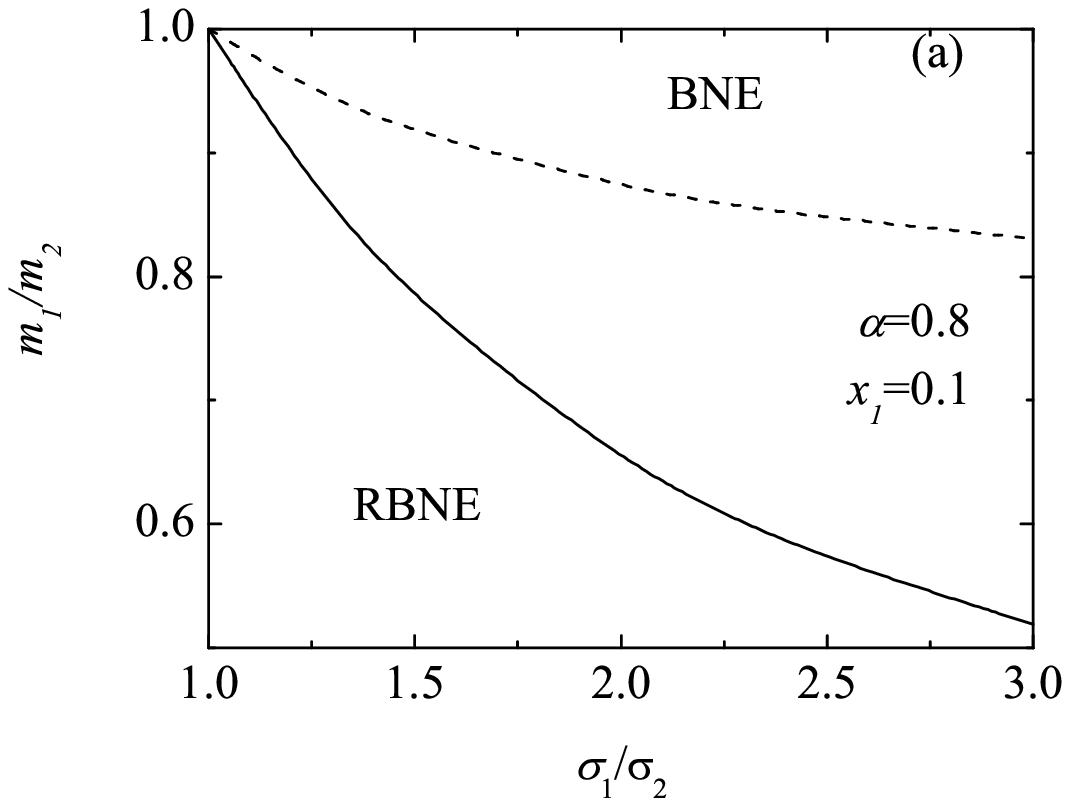}}&\resizebox{7.75cm}{!}
{\includegraphics{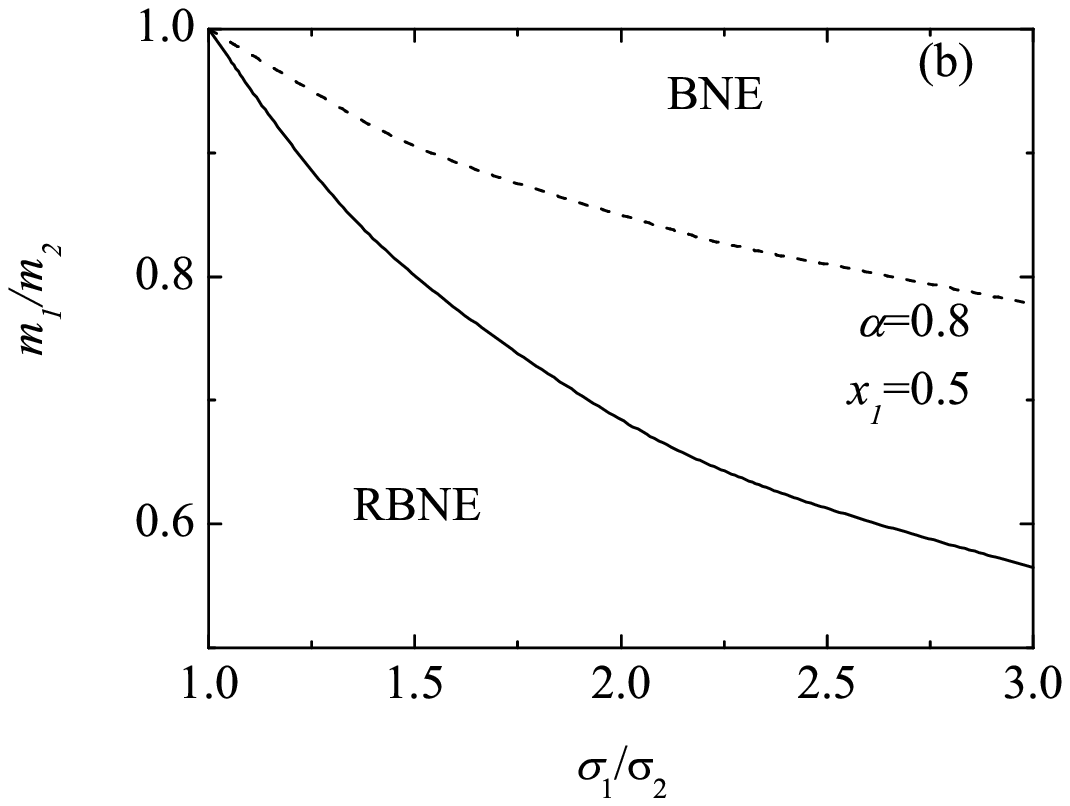}}
\end{tabular}
\end{center}
\caption{BNE/RBNE phase diagram for inelastic hard spheres at $\alpha=0.8$ and two different values of the mole fraction $x_1$: $x_1=0.1$ (panel (a)) and $x_1=0.5$ (panel (b)). Points above the curves correspond to $\Lambda >0$ (BNE) while points below the curves correspond to $\Lambda <0$ (RBNE). The dashed and solid lines are the results obtained from the first and second Sonine approximations, respectively.
\label{fig5}}
\end{figure*}
\begin{figure}[htbp]
\begin{center}
\resizebox{8cm}{!}{\includegraphics{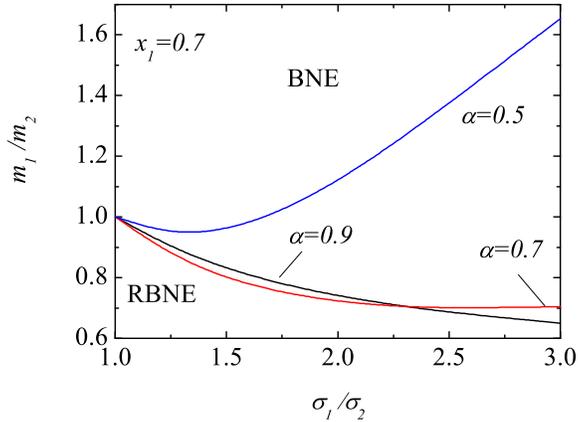}}
\end{center}
\caption{(color online) BNE/RBNE phase diagram for inelastic hard spheres with $x_1=0.7$ and three different values of the (common) coefficient of restitution $\alpha$. Points above the curves correspond to $\Lambda >0$ (BNE) while points below the curves correspond to $\Lambda <0$ (RBNE).} \label{fig6}
\end{figure}

Finally, we illustrate the form of the phase diagrams delineating the regimes between BNE and RBNE in the $\left( \sigma_1/\sigma_2, m_1/m_2 \right)$-plane. Figure \ref{fig5} shows phase diagrams for $\alpha=0.8$ and two values of the composition $x_1$. The first Sonine prediction is also shown for the sake of comparison. Although the first Sonine approximation shows the same trends of the phase diagram, it clearly overestimates the predictions of the second Sonine approximation, specially at large size ratios and small mass ratios. Regarding the influence of the concentration of the mixture $x_1$ on phase diagrams we observe that the BNE region is reduced as $x_1$ increases. On the other hand, this effect is less significant than for dense binary mixtures (see, for instance, Fig.\ 7 of Ref.\ \onlinecite{G11}). Moreover, in contrast to what
happens in the dense case, \cite{G11} at a given value of the concentration, the transition from BNE to RBNE may occur following two paths: i) along constant mass ratio $m_1/m_2$ with decreasing diameter ratio $\sigma_1/\sigma_2$, and ii) along constant diameter ratio $\sigma_1/\sigma_2$  with decreasing mass ratio $m_1/m_2$. Next, we study the impact of inelasticity on the form of the phase diagrams. Figure \ref{fig6} shows the phase diagram for $x_1=0.7$ and three values of $\alpha$ ($\alpha=$0.9, 0.7 and 0.5). The results show that the main effect of collisional dissipation is to reduce the size of the BNE region. This contrasts again with the results obtained from the first Sonine approximation for the Enskog equation (see, for instance, Fig.\ 5 of Ref.\ \onlinecite{G11}). The influence of dissipation on the BNE/RBNE phase diagram is much more significant for quite strong values of $\alpha$ (say for instance, $\alpha=0.5$) since the lines delineating the regimes between BNE and RBNE for $\alpha=0.9$ and 0.7 are quite similar (at least in the region of values of the diameter ratio explored).

\section{Summary and discussion}
\label{sec7}

In this paper we have determined the mass flux $\mathbf{j}_1^{(1)}$  of a granular binary mixture at low-density. The results have been obtained by solving the inelastic Boltzmann equation by means of the Chapman-Enskog method at the NS order. Three diffusion coefficients characterize the mass flux in the NS regime: the mutual diffusion coefficient $D$ (that couples $\mathbf{j}_1^{(1)}$ with the concentration gradient $\nabla x_1$), the pressure diffusion coefficient $D_p$ (that couples $\mathbf{j}_1^{(1)}$ with the pressure gradient $\nabla p$), and the thermal diffusion coefficient $D'$ (that couples $\mathbf{j}_1^{(1)}$ with the temperature gradient $\nabla T$). On the other hand, as for elastic collisions, \cite{CC70} the above coefficients [see Eqs.\ \eqref{3.2}--\eqref{3.4}] are defined in terms of quantities ${\boldsymbol {\cal A}}_1$, ${\boldsymbol {\cal B}}_1$, and ${\boldsymbol {\cal C}}_1$ which are the solutions of a set of linear integral equations [see Eqs.\ \eqref{3.5a}--\eqref{3.7b}]. Given that the above quantities cannot be \emph{exactly} obtained, they are approximated by a truncated Sonine polynomial expansion. This allows us to obtain explicit forms for the diffusion transport coefficients in terms of the coefficients of restitution and the parameters of the mixture (relative masses, diameters and concentration). Here, we have determined $D$, $D_p$ and $D'$ by considering two polynomials in the Sonine polynomial expansion [see Eqs.\ \eqref{3.15a}--\eqref{3.17b}]. This approximation is usually referred to as the second Sonine approximation. Our present study complements and extends previous works on diffusion transport coefficients carried out in the tracer limit. \cite{GM04,GF09}

As mentioned in the Introduction, previous results \cite{KCM87} derived many years ago for ordinary mixtures ($\alpha_{ij}=1$) have clearly shown the reliability of the second Sonine approximation for the mutual and thermal diffusion coefficients for a wide range of values of masses and sizes. These results have mainly encouraged the present work since the studies of the impact of the Sonine approximation on the NS transport coefficients are very scarce in the case of granular mixtures. On the other hand, given the technical difficulties involved in the evaluation of the second Sonine corrections to the transport coefficients, we have focussed on our efforts in the case of diffusion coefficients which are related to the lowest velocity moment (the mass flux) of the first-order distribution functions $f_i^{(1)}$.

In order to gauge the accuracy of the second Sonine approximation, we have compared our theoretical predictions for the mutual diffusion coefficient $D$ with numerical solutions of the Boltzmann equation by means of the DSMC. \cite{B94} Two situations have been considered: the self-diffusion problem (namely, when both species are mechanically equivalent) and the tracer limit (namely, when the concentration of one of the species is negligible). These are perhaps the two most simple situations where the coefficient $D$ can be measured from the mean square displacement of a tracer particle immersed in a granular gas under HCS. The simulations performed here consider more general situations than those analyzed in previous works \cite{GM04,GM07,GF09} where it was assumed that $\alpha_{22}=\alpha_{12}$. As in previous studies, the present comparison shows again that in general the second Sonine approximation to $D$ improves significantly the prediction of the first Sonine approximation, especially for high degree of dissipation and/or extreme mass or size ratios.

A second important issue covered in this paper has been the study of segregation and mixing of dissimilar grains. The understanding of physical mechanisms involved in segregation within polydisperse, rapid granular flows is perhaps one of the most important open challenges of granular gas research. Among the different mechanisms involved in segregation, thermal diffusion (segregation induced by a thermal gradient) becomes the most relevant one when the sample is vibrated at large shaking amplitude. In this regime, binary collisions prevail and the granular system behaves like a granular gas. In a steady state without shearing flows, the \emph{sign} of the thermal diffusion factor [defined by Eq.\ \eqref{6.1}] provides information on the tendency of each species to move towards the colder or hotter plate. The knowledge of the three diffusion transport coefficients allows one to compute the thermal diffusion factor $\Lambda$ in terms of the coefficients of restitution, the concentration and the mass and size ratios. The evaluation of the thermal diffusion factor is of central interest in the field of granular matter mainly due to its practical and industrial importance.

The analysis carried out here for segregation provides an extension of previous studies performed \cite{GF09} in the tracer limit ($x_1 \to 0$). Our present results show that the influence of collisional dissipation on thermal diffusion is in general important. This is clearly illustrated in Figs.\ \ref{vicente1} and \ref{vicente2} where the form of the inelastic thermal diffusion factor $\Lambda(\alpha)$ differs significantly from its elastic counterpart $\Lambda(1)$ even at moderate dissipation. Moreover, our study also reveals that the effect of the concentration $x_1$ on BNE-RBNE phase diagrams is less important than the one previously obtained for dense binary mixtures. \cite{G11} We expect the segregation criteria obtained here by using the second Sonine approximation can be tested against DSMC results, molecular dynamics (MD) simulations, and eventually experiments in real problems. We are currently working on DSMC and MD simulations adapted to the problem of segregation.

One of the main limitations of the present study is its restriction to dilute gases. Given that most of the experiments are carried out for dense granular systems, it would be convenient to extend the present results to densities beyond the low-density limit. The NS transport coefficients for  granular mixtures at \emph{moderate} densities (solid volume fractions typically smaller than or equal to 0.25) have been recently obtained from the Enskog kinetic equation \cite{GDH07} by considering the first Sonine approximation. The evaluation of the second Sonine expressions of the diffusion transport coefficients from the Enskog equation could be a possible future work. This extension could allow us to compare our theoretical results (based on the second Sonine approximation) with MD simulations performed at finite densities.

\acknowledgments

V.G. and F.V. acknowledge the support of the Spanish Government through Grants No. FIS2010-16587 and No. MAT2009-14351-C02-02 (F.V.). The first Grant has been partially financed by FEDER funds and by the Junta de Extremadura (Spain) through Grant No. GR10158. J. A. M. is grateful for the funding support provided by the Department of Energy (DE-FC26-07NT43098) and the National Science Foundation (CBET-0318999).

\appendix
\section{First and second Sonine approximations to the mass flux}
\label{appA}

In this Appendix we determine the first and second Sonine approximations to the diffusion coefficients $D$, $D_p$, and $D'$. These coefficients are defined by Eqs.\ \eqref{3.2}--\eqref{3.4} where the functions ${\boldsymbol {\cal A}}_{i}$, ${\boldsymbol {\cal B}}_{i}$, and ${\boldsymbol {\cal C}}_{i}$ are given by Eqs.\ \eqref{3.15a}--\eqref{3.17b}. Our goal is to evaluate the nine independent Sonine coefficients
\begin{equation}
\label{a1}
\{a_{1,1}; b_{1,1}; c_{1,1}; a_{1,2}; a_{2,2}; b_{1,2}; b_{2,2}; c_{1,2}; c_{2,2} \}.
\end{equation}
The three first coefficients ($a_{1,1}$, $b_{1,1}$, and $c_{1,1}$) are directly related to the diffusion coefficients $D$, $D_p$, and $D'$, respectively.

Substitution of Eqs.\ \eqref{3.15a}--\eqref{3.17b} into the integral equations \eqref{3.5a}, \eqref{3.6a}, and \eqref{3.7a} gives
\begin{eqnarray}
& & -\zeta ^{(0)}\left( T\partial _{T}+p\partial _{p}\right) \left[a_{1,1}{\bf V}_{1} +  a_{1,2}
{\bf S}_1({\bf V}_1)\right]f_{1,M}+a_{1,1}\left[ {\cal L}_{1}f_{1,M}{\bf V}_{1}-\delta \gamma {\cal M}
_{1}f_{2,M}{\bf V}_{2}\right]\nonumber\\
& &+ a_{1,2}{\cal L}_{1}f_{1,M}{\bf S}_{1}({\bf V}_1)+a_{2,2}{\cal M}
_{1}f_{2,M}{\bf S}_{2}({\bf V}_2)={\bf A}_{12},
\label{a2}
\end{eqnarray}
\begin{eqnarray}
& & \left[-\zeta ^{(0)}\left( T\partial _{T}+p\partial _{p}\right)-2\zeta^{(0)}\right]\left[b_{1,1}{\bf V}_{1} +  b_{1,2} {\bf S}_1({\bf V}_1)\right]f_{1,M}+b_{1,1}\left[ {\cal L}_{1}f_{1,M}{\bf V}_{1}-\delta \gamma {\cal M}
_{1}f_{2,M}{\bf V}_{2}\right]\nonumber\\
& &+ b_{1,2}{\cal L}_{1}f_{1,M}{\bf S}_{1}({\bf V}_1)+b_{2,2}{\cal M}
_{1}f_{2,M}{\bf S}_{2}({\bf V}_2)
={\bf B}_{12},
\label{a3}
\end{eqnarray}
\begin{eqnarray}
& & \left[-\zeta ^{(0)}\left( T\partial _{T}+p\partial _{p}\right)-\frac{1}{2}\zeta^{(0)}\right]\left[c_{1,1}{\bf V}_{1} +  c_{1,2}
{\bf S}_1({\bf V}_1)\right]f_{1,M}+c_{1,1}\left[ {\cal L}_{1}f_{1,M}{\bf V}_{1}-\delta \gamma {\cal M}
_{1}f_{2,M}{\bf V}_{2}\right]\nonumber\\
& & +c_{1,2}{\cal L}_{1}f_{1,M}{\bf S}_{1}({\bf V}_1)+c_{2,2}{\cal M}
_{1}f_{2,M}{\bf S}_{2}({\bf V}_2)
={\bf C}_{12},
\label{a4}
\end{eqnarray}
where
\begin{equation}
\label{a5}
{\bf A}_{12}={\bf A}_1+\left( \frac{\partial \zeta ^{(0)}}{\partial x_{1}}\right)
_{p,T}f_{1,M}\left[ p\left(b_{1,1}{\bf V}+b_{1,2}{\bf S}_1\right)+T\left(c_{1,1}{\bf V}+c_{1,2}
\mathbf {S}_1\right)\right],
\end{equation}
\begin{equation}
\label{a6}
{\bf B}_{12}={\bf B}_1+\frac{T\zeta^{(0)}}{p}f_{1,M}\left(c_{1,1}{\bf V}+c_{1,2}{\bf S}_1\right),
\end{equation}
\begin{equation}
\label{a7}
{\bf C}_{12}={\bf C}_1-\frac{p\zeta^{(0)}}{2T}f_{1,M}\left(b_{1,1}{\bf V}+b_{1,2}{\bf S}_1\right).
\end{equation}
Here, $\textbf{A}_1$, $\textbf{B}_1$, and $\textbf{C}_1$ are given by Eqs.\ \eqref{3.11}--\eqref{3.13}, respectively.
The corresponding counterparts of Eqs.\ (\ref{a2})--(\ref{a4}) can be obtained from them by just making the change $1\leftrightarrow 2$. Next, we multiply Eqs.\ (\ref{a2})--(\ref{a3}) by $m_1 {\bf V}_1$ and integrates over the velocity. The result is
\begin{widetext}
\begin{equation}
\left[ -\zeta ^{(0)}\left( T\partial _{T}+p\partial _{p}\right) +\nu \right]
n_{1}T_{1}a_{1,1}+n_1T_1\left(\tau_{11}a_{1,2}+\tau_{12}a_{2,2}\right)=-\left( \frac{
\partial }{\partial x_{1}}n_{1}T_{1}\right) _{p,T}
+\left( \frac{\partial \zeta ^{(0)}}{\partial x_{1}}\right)
_{p,T}n_{1}T_{1}\left( pb_{1,1}+Tc_{1,1}\right) ,  \label{a8}
\end{equation}
\begin{equation}
\left[ -\zeta ^{(0)}\left( T\partial _{T}+p\partial _{p}\right) -2\zeta
^{(0)}+\nu \right] n_{1}T_{1}b_{1,1}+n_1T_1\left(\tau_{11}b_{1,2}+\tau_{12}b_{2,2}\right)
=-\frac{n_{1}T_{1}}{p}\left( 1-\frac{m_{1}nT}{\rho T_{1}}\right)+\frac{T\zeta ^{(0)}}{p}n_{1}T_{1}c_{1,1},  \label{a9}
\end{equation}
\begin{equation}
\left[-\zeta ^{(0)}\left( T\partial _{T}+p\partial _{p}\right) -\frac{1}{2}
\zeta ^{(0)}+\nu \right] n_{1}T_{1}c_{1,1}+n_1T_1\left(\tau_{11}b_{1,2}+\tau_{12}b_{2,2}\right)=-\frac{p\zeta ^{(0)}}{2T}n_{1}T_{1}b_{1,1}.
\label{a10}
\end{equation}
\end{widetext}
Here, we have introduced the collision frequencies
\begin{eqnarray}
\nu  &=&\frac{1}{dn_{1}T_{1}}\int d{\bf V}_{1}m_{1}{\bf V}_{1}\cdot \left[
{\cal L}_{1}f_{1,M}{\bf V}_{1}-\delta \gamma {\cal M}_{1}f_{2,M}{\bf V}_{2}
\right]   \nonumber \\
&=&-\frac{1}{dn_{1}T_{1}}\int d{\bf V}_{1}m_{1}{\bf V}_{1}\cdot \left(
J_{12}[{\bf v}_{1}|f_{1,M}{\bf V}_{1},f_{2}^{(0)}]-\delta \gamma J_{12}[{\bf
v}_{1}|f_{1}^{(0)},f_{2,M}{\bf V}_{2}]\right),  \label{a11}
\end{eqnarray}
\begin{equation}
\label{a12}
\tau_{ii}=\frac{1}{dn_iT_i}\int d{\bf v}_1 m_i {\bf V}_1
\cdot {\cal L}_i\left(f_{i,M}{\bf S}_i\right),
\end{equation}
\begin{equation}
\label{a13}
\tau_{ij}=\frac{1}{dn_iT_i}\int d{\bf v}_1 m_i {\bf V}_1
\cdot {\cal M}_i\left(f_{j,M}{\bf S}_j\right), \quad i\neq j.
\end{equation}
┐From dimensional analysis, $n_{1}T_{1}a_{1,1}\sim T^{1/2}$, $
n_{1}T_{1}b_{1,1}\sim T^{1/2}/p$, and $n_{1}T_{1}c_{1,1}\sim T^{-1/2}$. Thus, the
temperature derivatives can be performed in Eqs.\ (\ref{a8})--(\ref{a10})
and the result is
\begin{equation}
\left(\nu -\frac{1}{2}\zeta ^{(0)}\right) a_{1,1}+\tau_{11}a_{1,2}+\tau_{12}a_{2,2}=
-\left( \frac{\partial }{
\partial x_{1}}\ln n_{1}T_{1}\right) _{p,T}
+\left( \frac{\partial \zeta^{(0)}}{\partial x_{1}}\right) _{p,T}\left( pb_{1,1}+Tc_{1,1}\right),
\label{a14}
\end{equation}
\begin{equation}
\left(\nu -\frac{3}{2}\zeta ^{(0)}\right) b_{1,1}+\tau_{11}b_{1,2}+\tau_{12}b_{2,2}=
-\frac{1}{p}\left( 1-\frac{m_{1}n T}{\rho T_{1}}\right) +\frac{T\zeta ^{(0)}}{p}c_{1,1},
\label{a15}
\end{equation}
\begin{equation}
\nu c_{1,1}+\tau_{11}c_{1,2}+\tau_{12}c_{2,2}=-\frac{p\zeta ^{(0)}}{2T}b_{1,1}.  \label{a16}
\end{equation}

If only the first Sonine approximation is considered (which means $a_{i,2}=b_{i,2}=c_{i,2}=0$), the solution to Eqs.\ (\ref{a14})--(\ref{a16}) is
\begin{equation}
a_{1,1}[1]=-\left(\nu -\frac{1}{2}\zeta ^{(0)}\right) ^{-1}\left[ \left(\frac{
\partial }{\partial x_{1}}\ln n_{1}T_{1}\right) _{p,T}- \left( \frac{\partial
\zeta ^{(0)}}{\partial x_{1}}\right) _{p,T}\left( p\;b_{1,1}[1]+T\;c_{1,1}[1]\right)
\right] ,   \label{a17}
\end{equation}
\begin{equation}
b_{1,1}[1]=-\frac{1}{p}\left( 1-\frac{m_{1}nT}{\rho T_{1}}\right) \left( \nu -
\frac{3}{2}\zeta ^{(0)}+\frac{\zeta ^{(0)2}}{2\nu }\right)^{-1},
\label{a18}
\end{equation}
\begin{equation}
c_{1,1}[1]=-\frac{p\zeta ^{(0)}}{2T\nu }b_{1,1}[1],  \label{a19}
\end{equation}
Here, $a_{i,2}[1]$, $b_{i,2}[1]$, and $c_{i,2}[1]$ denotes the first Sonine approximation to $a_{i,2}$, $b_{i,2}$, and $c_{i,2}$, respectively. From Eqs.\ \eqref{a17}--\eqref{a19} one gets the first Sonine expressions \eqref{3.25}--\eqref{3.27} for $D$, $D_p$ and $D'$, respectively.

To close the problem, we multiply now Eqs.\ (\ref{a2})--(\ref{a4}) by ${\bf S}_1({\bf V}_1)$ and integrates over the velocity. Following identical mathematical steps as before and after some algebra one gets
\begin{equation}
\left( \nu_{11} -\frac{3}{2}\zeta ^{(0)}\right) a_{1,2}+\nu_{12}a_{2,2}-\left( \frac{\partial \zeta
^{(0)}}{\partial x_{1}}\right) _{p,T}\left( pb_{1,2}+Tc_{1,2}\right)=-\left(\frac{\zeta^{(0)}}{T_1}-\Delta_{12}\right)a_{1,1}
-\frac{1}{2}\frac{T^2}{T_1^3}\left(\frac{\partial \gamma_1^2}{\partial x_1}\right)_{p,T},
\label{a20}
\end{equation}
\begin{equation}
\label{a21}
\left( \nu_{11} -\frac{5}{2}\zeta ^{(0)}\right) b_{1,2}+\nu_{12}b_{2,2}-\frac{T\zeta^{(0)}}{p}c_{1,2}
-\left(\frac{\zeta^{(0)}}{T_1}-\Delta_{12}\right)b_{1,1}=0,
\end{equation}
\begin{equation}
\label{a22}
\left( \nu_{11} -\zeta ^{(0)}\right) c_{1,2}+\nu_{12}c_{2,2}+\frac{p\zeta^{(0)}}{2T}b_{1,2}
-\left(\frac{\zeta^{(0)}}{T_1}-\Delta_{12}\right)c_{1,1}=-\frac{1}{TT_1}.
\end{equation}
Here, $\Delta_{12}=\lambda_{11}-\delta \gamma \lambda_{12}$, where $\delta\equiv x_1/x_2$, $x_2=1-x_1$, $\gamma\equiv T_1/T_2$ and
\begin{equation}
\label{a23}
\lambda_{ii}=\frac{2}{d(d+2)}\frac{m_i}{n_iT_i^3}\int d{\bf v}_1 {\bf S}_i
\cdot {\cal L}_i\left(f_{i,M}{\bf V}_1\right),
\end{equation}
\begin{equation}
\label{a24}
\lambda_{ij}=\frac{2}{d(d+2)}\frac{m_i}{n_iT_i^3}\int d{\bf v}_1 {\bf S}_i
\cdot {\cal M}_i\left(f_{j,M}{\bf V}_2\right), \quad i\neq j.
\end{equation}
In addition, in Eqs.\ \eqref{a20}--\eqref{a22}, we have introduced the collision frequencies
\begin{equation}
\label{a25}
\nu_{ii}=\frac{2}{d(d+2)}\frac{m_i}{n_iT_i^3}\int d{\bf v}_1 {\bf S}_i
\cdot {\cal L}_i\left(f_{i,M}{\bf S}_i\right),
\end{equation}
\begin{equation}
\label{a26}
\nu_{ij}=\frac{2}{d(d+2)}\frac{m_i}{n_iT_i^3}\int d{\bf v}_1 {\bf S}_i
\cdot {\cal M}_i\left(f_{j,M}{\bf S}_j\right), \quad i\neq j.
\end{equation}
The corresponding integral equations verifying the remaining coefficients $a_{2,2}$, $b_{2,2}$, and $c_{2,2}$ can be obtained from Eqs.\ \eqref{a20}--\eqref{a22}, respectively, by interchanging $1\leftrightarrow 2$. Note that upon writing Eqs.\ \eqref{a20}--\eqref{a22} we have neglected the non-Gaussian corrections to $f_i^{(0)}$.

Equations \eqref{a14}--\eqref{a16} along with Eqs.\ \eqref{a20}--\eqref{a22} can be written in a more compact form by using matrix notation. For the sake of convenience, let us introduce the dimensionless coefficients $a_{1,1}^*=\nu_0 a_{1,1}$, $b_{1,1}^*=p\nu_0 b_{1,1}$, $c_{1,1}^*=T\nu_0 c_{1,1}$, $a_{i,2}^*=T\nu_0 a_{i,2}$, $b_{i,2}^*=pT\nu_0 b_{i,2}$, and $c_{i,2}^*=T^2\nu_0 c_{i,2}$, where $\nu_0$ is defined by Eq.\ \eqref{3.24}. Let us introduce the column matrix $\mathsf{X}$ by
\begin{equation}
\label{a27}
\{a_{1,1}^*; b_{1,1}^*; c_{1,1}^*; a_{1,2}^*; a_{2,2}^*; b_{1,2}^*; b_{2,2}^*; c_{1,2}^*; c_{2,2}^*\}.
\end{equation}
Therefore, according to Eqs.\ \eqref{a14}--\eqref{a16} and \eqref{a20}--\eqref{a22}, the coupled set of nine equations for the unknowns can be rewritten in matrix form as
\begin{equation}
\label{a28}
\Omega_{\sigma \sigma'}X_{\sigma'}=Y_{\sigma},
\end{equation}
where the square matrix $\mathsf{\Omega}$ is
\begin{equation}
\label{a28bis}
\mathsf{\Omega}=\mathsf{\Omega}^{(0)}+\mathsf{\Omega}^{(1)},
\end{equation}
\begin{widetext}
\begin{equation}
\label{a29}
\mathsf{\Omega}^{(0)}=\left(
\begin{array} {ccccccccc}
\nu^*-\frac{1}{2}\zeta^*&0&0&\tau_{11}^*&\tau_{12}^*&0&0&0&0\\
0&\nu^*-\frac{3}{2}\zeta^{*}&0&0&0&\tau_{11}^*&\tau_{12}^*&0&0\\
0&0&\nu^*&0&0&0&0&\tau_{11}^*&\tau_{12}^*\\
0&0&0&\nu_{11}^*-\frac{3}{2}\zeta^{*}&\nu_{12}^*&0&0&0&0\\
0&0&0&\nu_{21}^*&\nu_{22}^*-\frac{3}{2}\zeta^{*}&0&0&0&0\\
0&0&0&0&0&\nu_{11}^*-\frac{5}{2}\zeta^{*}&\nu_{12}^*&0&0\\
0&0&0&0&0&\nu_{21}^*&\nu_{22}^*-\frac{5}{2}\zeta^{*}&0&0\\
0&0&0&0&0&0&0&\nu_{11}^*-\zeta^{*}&\nu_{12}^*\\
0&0&0&0&0&0&0&\nu_{21}^*&\nu_{22}^*-\zeta^{*}
\end{array}
\right),
\end{equation}
\begin{equation}
\label{a30}
\mathsf{\Omega}^{(1)}=\left(
\begin{array} {ccccccccc}
0&-\left(\frac{\partial \zeta ^{*}}{\partial x_{1}}\right)_{p,T}&-\left(\frac{\partial \zeta ^{*}}{\partial x_{1}}\right)_{p,T}&0&0&0&0&0&0\\
0&0&-\zeta ^{*}&0&0&0&0&0&0\\
0&\zeta ^{*}/2&0&0&0&0&0&0&0\\
\omega _{12}^*&0&0&0&0&
-\left(\frac{\partial \zeta ^{*}}{\partial x_{1}}\right)_{p,T}&0&
-\left( \frac{\partial \zeta ^{*}}{\partial x_{1}}\right)_{p,T}&0 \\
\omega _{21}^*&0&0&
0&0&0&
-\left( \frac{\partial \zeta ^{*}}{\partial x_{1}}\right)_{p,T}&0&
-\left( \frac{\partial \zeta ^{*}}{\partial x_{1}}\right)_{p,T}\\
0&\omega _{12}^*&0&
0& 0&0&0& -\zeta^{*}&0\\
0&\omega _{21}^*&0&
0& 0 & 0& 0 & 0 &
-\zeta^{*}\\
0&0&\omega _{12}^*&
0& 0& \zeta^{*}/2&0&0& 0\\
0&0&\omega _{21}^*&
0& 0& 0&\zeta^{*}/2&0&0
\end{array}
\right).
\end{equation}
The column matrix $\mathsf{Y}$ is given by
\begin{equation}
\label{a31}
\mathsf{Y}=\left(
\begin{array}{c}
-\frac{1}{x_1\gamma_1}\left(\frac{\partial }{\partial x_{1}}x_{1}\gamma_{1}\right) _{p,T}\\
-\left(1-\frac{\mu(1+\delta)}{\gamma_1(1+\mu\delta)}\right)\\
0\\
-\frac{1}{2\gamma_1^3}\left(\frac{\partial \gamma_1^2}{\partial x_1}\right)_{p,T}\\
-\frac{1}{2\gamma_2^3}\left(\frac{\partial \gamma_2^2}{\partial x_1}\right)_{p,T}\\
0\\
0\\
-\gamma_1^{-1}\\
-\gamma_2^{-1}
\end{array}
\right).
\end{equation}
\end{widetext}
In the above equations, we have introduced the reduced quantities $\nu^*=\nu/\nu_0$, $\tau_{ij}^*=\tau_{ij}/T\nu_0$, $\nu_{ij}^*=\nu_{ij}/\nu_0$, and
\begin{equation}
\label{a32}
\omega_{12}^*=\Delta_{12}^*-\frac{\zeta^{*}}{\gamma_1},\quad \omega_{21}^*=-\delta \gamma \left(\Delta_{21}^*-\frac{\zeta^{*}}{\gamma_2}\right),\quad
\Delta_{ij}^*=\frac{T \Delta_{ij}}{\nu_0}.
\end{equation}

The solution to Eq.\ \eqref{a28} is
\begin{equation}
\label{a33}
X_{\sigma}=(\Omega^{-1})_{\sigma \sigma'}Y_{\sigma'}.
\end{equation}
┐From this relation one gets the second Sonine corrections to the coefficients $a_{11}$, $b_{11}$, and $c_{11}$.

\section{Reduced collision frequencies and cooling rates}
\label{appB}

In this Appendix we provide the explicit expressions of the (reduced) collision frequencies needed to evaluate $D[2]$, $D_p[2]$, and $D'[2]$. As said in the main text, to evaluate them we take the local Maxwellian approximations \eqref{3.14} for the zeroth-order distributions $f_i^{(0)}$. These collision frequencies have been already evaluated in the $d$ dimensional case. \cite{GM07,GF09} They are given by
\begin{equation}
\label{b1}
\nu^*=\frac{2\pi^{\frac{d}{2}-1}}{d\Gamma\left(\frac{d}{2}\right)}
(1+\alpha_{12})
\left(\frac{\theta_1+\theta_2}{\theta_1\theta_2}\right)^{1/2}\left(x_2 \mu_{21}+
x_1 \mu_{12}\right),
\end{equation}
\begin{equation}
\label{b2}
\tau_{11}^*=\frac{\pi^{\frac{d}{2}-1}}{d\Gamma\left(\frac{d}{2}\right)}x_2
(1+\alpha_{12})\frac{\theta_2^{1/2}(\theta_1+\theta_2)^{-1/2}}{\theta_1^{3/2}},
\end{equation}
\begin{equation}
\label{b3}
\tau_{12}^*=-\frac{\pi^{\frac{d}{2}-1}}{d\Gamma\left(\frac{d}{2}\right)}
x_2(1+\alpha_{12})\gamma^{-1}\frac{\theta_1^{1/2}(\theta_1+\theta_2)^{-1/2}}{\theta_2^{3/2}},
\end{equation}
\begin{eqnarray}
\label{b4}
\Delta_{12}^*&=&\frac{\pi^{\frac{d}{2}-1}}{\Gamma\left(\frac{d}{2}\right)}\frac{\sqrt{2}}{d}
x_1\left(\frac{\sigma_1}{\sigma_{12}}\right)^{d-1}\sqrt{\frac{\mu_{21}}{\gamma_1}}(1-\alpha_{11}^2)
\nonumber\\
& & +\frac{\pi^{\frac{d}{2}-1}}{\Gamma\left(\frac{d}{2}\right)}\frac{2}{d(d+2)}
\mu_{21}^{-1}\gamma_1^{-3}(1+\alpha_{12})\left(\theta_1+\theta_2\right)^{-1/2}\left(\theta_1\theta_2\right)^{-3/2}
\left(x_2 A-\gamma x_1 B\right),
\end{eqnarray}
\begin{eqnarray}
\label{b5}
\nu_{11}^*&=&\frac{\pi^{\frac{d}{2}-1}}{\Gamma\left(\frac{d}{2}\right)}
\frac{8}{d(d+2)}\left(\frac{\sigma_1}{\sigma_{12}}\right)^{d-1}
x_1 (2\theta_1)^{-1/2}
(1+\alpha_{11})\left[\frac{d-1}{2}+\frac{3}{16}(d+8)(1-\alpha_{11})\right]\nonumber\\
& & +\frac{\pi^{(d-1)/2}}{\Gamma\left(\frac{d}{2}\right)}
\frac{1}{d(d+2)}x_2\mu_{21}(1+\alpha_{12})\left(\frac{\theta_1}
{\theta_2(\theta_1+\theta_2)}\right)^{3/2}\left[E-(d+2)\frac{\theta_1+\theta_2}{\theta_1}A\right],
\end{eqnarray}
\begin{equation}
\label{b6}
\nu_{12}^*=-\frac{\pi^{\frac{d}{2}-1}}{\Gamma\left(\frac{d}{2}\right)}
\frac{1}{d(d+2)}x_2\frac{\mu_{21}^2}{\mu_{12}}(1+\alpha_{12})\left(\frac{\theta_1}
{\theta_2(\theta_1+\theta_2)}\right)^{3/2}\left[F+(d+2)\frac{\theta_1+\theta_2}{\theta_2}B\right].
\end{equation}
In the above equations, $\mu_{ij}=m_i/(m_i+m_j)$, and
\begin{equation}
\label{b7} \theta_i=\frac{m_i}{\gamma_i}\sum_{j=1}^2\,m_j^{-1}.
\end{equation}
In addition, the quantities $A$, $B$, $E$, and $F$ are given, respectively, as
\begin{eqnarray}
\label{b8}
A&=&(d+2)(2\beta_{12}+\theta_2)+\mu_{21}(\theta_1+\theta_2)\left\{(d+2)(1-\alpha_{12})
-[(11+d)\alpha_{12}-5d-7]\beta_{12}\theta_1^{-1}\right\}\nonumber\\
& &
+3(d+3)\beta_{12}^2\theta_1^{-1}+2\mu_{21}^2\left(2\alpha_{12}^{2}-\frac{d+3}{2}\alpha
_{12}+d+1\right)\theta_1^{-1}(\theta_1+\theta_2)^2-
(d+2)\theta_2\theta_1^{-1}(\theta_1+\theta_2), \nonumber\\
\end{eqnarray}
\begin{eqnarray}
\label{b9} B&=&
(d+2)(2\beta_{12}-\theta_1)+\mu_{21}(\theta_1+\theta_2)\left\{(d+2)(1-\alpha_{12})
+[(11+d)\alpha_{12}-5d-7]\beta_{12}\theta_2^{-1}\right\}\nonumber\\
& &
-3(d+3)\beta_{12}^2\theta_2^{-1}-2\mu_{21}^2\left(2\alpha_{12}^{2}-\frac{d+3}{2}\alpha
_{12}+d+1\right)\theta_2^{-1}(\theta_1+\theta_2)^2+
(d+2)(\theta_1+\theta_2), \nonumber\\
\end{eqnarray}
\begin{eqnarray}
\label{b10} E&=&
 2\mu_{21}^2\theta_1^{-2}(\theta_1+\theta_2)^2
\left(2\alpha_{12}^{2}-\frac{d+3}{2}\alpha_{12}+d+1\right)
\left[(d+2)\theta_1+(d+5)\theta_2\right]\nonumber\\
& & -\mu_{21}(\theta_1+\theta_2)
\left\{\beta_{12}\theta_1^{-2}[(d+2)\theta_1+(d+5)\theta_2][(11+d)\alpha_{12}
-5d-7]\right.\nonumber\\
& & \left.
-\theta_2\theta_1^{-1}[20+d(15-7\alpha_{12})+d^2(1-\alpha_{12})-28\alpha_{12}]
-(d+2)^2(1-\alpha_{12})\right\}
\nonumber\\
& & +3(d+3)\beta_{12}^2\theta_1^{-2}[(d+2)\theta_1+(d+5)\theta_2]+
2\beta_{12}\theta_1^{-1}[(d+2)^2\theta_1+(24+11d+d^2)\theta_2]
\nonumber\\
& & +(d+2)\theta_2\theta_1^{-1}
[(d+8)\theta_1+(d+3)\theta_2]-(d+2)(\theta_1+\theta_2)\theta_1^{-2}\theta_2
[(d+2)\theta_1+(d+3)\theta_2],\nonumber\\
\end{eqnarray}
\begin{eqnarray}
\label{b11} F&=&
 2\mu_{21}^2\theta_2^{-2}(\theta_1+\theta_2)^2
\left(2\alpha_{12}^{2}-\frac{d+3}{2}\alpha_{12}+d+1\right)
\left[(d+5)\theta_1+(d+2)\theta_2\right]\nonumber\\
& & -\mu_{21}(\theta_1+\theta_2)
\left\{\beta_{12}\theta_2^{-2}[(d+5)\theta_1+(d+2)\theta_2][(11+d)\alpha_{12}
-5d-7]\right.\nonumber\\
& & \left.
+\theta_1\theta_2^{-1}[20+d(15-7\alpha_{12})+d^2(1-\alpha_{12})-28\alpha_{12}]
+(d+2)^2(1-\alpha_{12})\right\}
\nonumber\\
& & +3(d+3)\beta_{12}^2\theta_2^{-2}[(d+5)\theta_1+(d+2)\theta_2]-
2\beta_{12}\theta_2^{-1}[(24+11d+d^2)\theta_1+(d+2)^2\theta_2]
\nonumber\\
& & +(d+2)\theta_1\theta_2^{-1}
[(d+3)\theta_1+(d+8)\theta_2]-(d+2)(\theta_1+\theta_2)\theta_2^{-1}
[(d+3)\theta_1+(d+2)\theta_2]. \nonumber\\
\end{eqnarray}
Here, $\beta_{12}=\mu_{12}\theta_2-\mu_{21}\theta_1$. From Eqs.\
(\ref{b2})--(\ref{b6}), one easily gets the expressions for
$\tau_{22}^*$, $\tau_{21}^*$, $\Delta_{21}^*$, $\nu_{22}^*$ and $\nu_{21}^*$ by interchanging
$1\leftrightarrow 2$.

Finally, the temperature ratio $\gamma$ is determined from the condition \cite{GD99a}
\begin{equation}
\label{b12}
\zeta_1^*=\zeta_2^*=\zeta^*,
\end{equation}
where the dimensionless cooling rate is
\begin{eqnarray}
\label{b13}
\zeta_1^*&=& \frac{\sqrt{2}\pi^{\frac{d-1}{2}}}{d\Gamma\left(\frac{d}{2}\right)}
x_1\left(\frac{\sigma_1}{\sigma_{12}}\right)^{d-1}\theta_1^{-1/2}(1-\alpha_{11}^2)
\nonumber\\
& &+\frac{4\pi^{\frac{d-1}{2}}}{d\Gamma\left(\frac{d}{2}\right)}x_2\mu_{21}\left(\frac{\theta_1+
\theta_2}{\theta_1\theta_2}\right)^{1/2}(1+\alpha_{12})\left[1-\frac{\mu_{21}}{2}
(1+\alpha_{12})\frac{\theta_1+\theta_2}{\theta_2}\right].
\end{eqnarray}
The expression of $\zeta_2^*$ can be obtained form the change $1\leftrightarrow 2$. Once the temperature ratio $\gamma$ is known, the partial temperature ratios $\gamma_i=T_i/T$ $(i=1,2)$ can be expressed in terms of the (global) temperature as
\begin{equation}
\label{b14} \gamma_1=\frac{\gamma}{1+x_1(\gamma-1)},\quad
\gamma_2=\frac{1}{1+x_1(\gamma-1)}.
\end{equation}


\begin{thebibliography} {99}

\bibitem{GS95}A. Goldshtein and M. Shapiro, ``Mechanics of collisional motion of granular materials.
Part 1. General hydrodynamic equations,'' J. Fluid Mech. {\bf 282}, 75 (1995).

\bibitem{BDS97}J. J. Brey, J. W. Dufty, and A. Santos, ``Dissipative
dynamics for hard spheres,'' J. Stat. Phys. {\bf 87}, 1051 (1997).

\bibitem{BP04}N. V. Brilliantov and T. P\"oschel, {\em Kinetic Theory
of Granular Gases} (Oxford University Press, Oxford, 2004).

\bibitem{CC70}S. Chapman and T. G. Cowling, {\em The Mathematical Theory of Nonuniform Gases}
(Cambridge University Press, Cambridge, 1970).

\bibitem{BRC99}J. J. Brey, M. J. Ruiz-Montero, and D. Cubero, ``On the validity of linear hydrodynamics for low-density granular flows described by the Boltzmann equation,'' Europhys. Lett. \textbf{48}, 359 (1999).

\bibitem{GM02}V. Garz\'o and J. M. Montanero, ``Transport coefficients of a heated granular gas,'' Physica A
{\bf 313}, 336 (2002).

\bibitem{BR04}J. J. Brey and M. J. Ruiz-Montero, ``Simulation
study of the Green-Kubo relations for dilute granular gases,'' Phys. Rev. E {\bf 70}, 051301 (2004).

\bibitem{BDKS98}J. J. Brey, J. W. Dufty, C. S. Kim, and A. Santos,
``Hydrodynamics for granular flow at low-density,'' Phys. Rev. E {\bf 58}, 4638 (1998).

\bibitem{GD99}V. Garz\'{o} and J. W. Dufty, ``Dense fluid transport for inelastic hard spheres,''
Phys. Rev. E {\bf 59}, 5895 (1999).


\bibitem{L05}J. F. Lutsko, ``Transport properties of dense dissipative hard-sphere fluids for arbitrary energy loss models, '' Phys. Rev. E {\bf 72}, 021306 (2005).


\bibitem{GSM07}V. Garz\'o, A. Santos, and J. M. Montanero, ``Modified Sonine approximation for the Navier-Stokes transport coefficients of a granular gas,┤┤ Physica A {\bf 376}, 94 (2007).


\bibitem{NBSG07}H. Noskowicz, O. Bar-Lev, D. Serero, and I. Goldhirsch, ``Computer-aided kinetic theory and granular gases,'' Europhys. Lett. \textbf{79}, 60001 (2007).

\bibitem{GD02}V. Garz\'o and J. W. Dufty, ``Hydrodynamics for a granular binary mixture
at low density,'' Phys. Fluids {\bf 14}, 1476 (2002).

\bibitem{JM89}J. T. Jenkins and F. Mancini, ``Kinetic theory for binary
mixtures of smooth, nearly elastic spheres,'' Phys. Fluids A {\bf 1}, 2050 (1989).

\bibitem{Z95}P. Zamankhan, ``Kinetic theory for
multicomponent dense mixtures of slightly inelastic spherical
particles,'' Phys. Rev. E {\bf 52}, 4877 (1995).

\bibitem{AW98}B. Arnarson and J. T. Willits, ``Thermal diffusion in binary mixtures
of smooth, nearly elastic spheres with and without gravity,'' Phys. Fluids {\bf 10}, 1324 (1998).

\bibitem{WA99}J. T. Willits and B. Arnarson, ``Kinetic theory of a binary mixture of nearly elastic
disks,'' Phys. Fluids {\bf 11}, 3116 (1999).

\bibitem{SGNT06}D. Serero, I. Goldhirsch, S. H. Noskowicz, and M.-L. Tan,
``Hydrodynamics of granular gases and granular gas mixtures,'' J. Fluid Mech. {\bf 554}, 237 (2006).

\bibitem{Serero09}
D. Serero, S. H. Noskowicz, M.-L. Tan, and I. Goldhirsch, ``Binary granular gas mixtures: Theory, layering effects, and some open questions,'' Eur. Phys. J. Special Topics \textbf{179}, 221 (2009).

\bibitem{B94}G. A. Bird, {\em Molecular Gas Dynamics and the Direct
Simulation Monte Carlo of Gas Flows} (Clarendon, Oxford, 1994).

\bibitem{GM04}V. Garz\'o and J. M. Montanero, ``Diffusion of impurities in a granular gas,'' Phys. Rev. E
{\bf 69}, 021301 (2004).


\bibitem{GM07}V. Garz\'o and J. M. Montanero, ``Navier-Stokes transport coefficients of d-dimensional granular binary mixtures at low density,'' J. Stat. Phys. {\bf 129}, 27 (2007).

\bibitem{MG03a}J. M. Montanero and V. Garz\'o, ``Shear viscosity for a heated granular binary mixture
at low-density,'' Phys. Rev. E {\bf 67}, 021308 (2003).


\bibitem{GDH07}V. Garz\'o, J. W. Dufty and C. M. Hrenya, ``Enskog theory for polydisperse granular mixtures. I. Navier-Stokes order transport,'' Phys. Rev. E {\bf 76}, 031303 (2007); V. Garz\'o, C. M. Hrenya and J. W. Dufty, Enskog theory for polydisperse granular mixtures. II. Sonine polynomial approximation,'' Phys. Rev. E {\bf 76}, 031304 (2007); J. A. Murray, V. Garz\'o and C. M. Hrenya,
     ``Enskog kinetic theory for polydisperse granular mixtures. III. Comparison of dense and dilute transport coefficients and equations of state for a binary mixture,'' Powder Tech. \textbf{220}, 24 (2012).

\bibitem{GM03b} V. Garz\'o and J. M. Montanero, ``Shear viscosity for a moderately dense granular binary mixture,'' Phys. Rev. E {\bf 68}, 041302 (2003).

\bibitem{GF09}V. Garz\'o and F. Vega Reyes, ``Mass transport of impurities in a moderately dense granular gas,'' Phys. Rev. E \textbf{79}, 041303 (2009); ``Segregation of an intruder in a heated granular gas,'' Phys. Rev. E \textbf{85}, 021308 (2012).



\bibitem{GMD06}V. Garz\'o, J. M. Montanero, and J. W. Dufty, ``Mass
and heat fluxes for a binary granular mixture at low-density,'' Phys. Fluids {\bf 18}, 083305 (2006).

\bibitem{MG02}J. M. Montanero and V. Garz\'o, ``Monte Carlo simulation of the homogeneous cooling state for a granular mixture,'' Gran. Matt. {\bf 4}, 17 (2002).

\bibitem{simulations}See for instance,  A. Barrat and E. Trizac, ``Lack of energy equipartition in homogeneous heated binary granular mixtures,'' Gran. Matt. {\bf 4}, 57  (2002); S. R. Dahl, C. M. Hrenya, V. Garz\'o, and J. W. Dufty, ``Kinetic temperatures
for a granular mixture,'' Phys. Rev. E {\bf 66}, 041301 (2002); R. Pagnani, U. M. B. Marconi, and A. Puglisi,
``Driven low density granular mixtures,'' {\em ibid.} {\bf 66}, 051304 (2002); D. Paolotti, C. Cattuto, U. M. B.
Marconi, and A. Puglisi, ``Dynamical properties of vibrofluidized granular mixtures,'' Gran. Matt. {\bf
5}, 75 (2003); P. Krouskop and J. Talbot, ``Mass and size effects in three-dimensional vibrofluidized granular
mixtures,'' Phys. Rev. E {\bf 68}, 021304 (2003); H. Wang, G. Jin, and Y. Ma, ``Simulation study on kinetic
temperatures of vibrated binary granular mixtures,'' {\em ibid.} {\bf 68}, 031301 (2003);  M. Schr\"oter, S. Ulrich , J. Kreft , J. B. Swift, and H. L. Swinney, ``Mechanisms in the size segregation of a binary granular mixture,'' {\em ibid.} {\bf 74}, 011307 (2006).


\bibitem{exp}R. D. Wildman and D. J. Parker, ``Coexistence of two granular
temperatures in binary vibrofluidized beds,'' Phys. Rev. Lett. {\bf 88}, 064301 (2002); K. Feitosa and N.
Menon, ``Breakdown of energy equipartition in a 2D binary vibrated granular gas,'' {\em ibid.} {\bf 88}, 198301
(2002).


\bibitem{KCM87}J. Kincaid, E. G. D. Cohen and M. L\'opez de Haro, ``The Enskog theory for multicomponent mixtures: IV. Thermal diffusion,'' J. Chem. Phys. {\bf 86}, 963 (1987).


\bibitem{Peter11}P. P. Mitrano, S. R. Dahl, D. J. Cromer, M. S. Pacella, and C. M. Hrenya, ``Instabilities in the homogeneous cooling of a granular gas: A quantitative assessment of kinetic-theory predictions,'' Phys. Fluids \textbf{23}, 093303 (2011); P. P. Mitrano, V. Garz\'o, A. M. Hilger, C. J. Ewasko, and C. M. Hrenya, ``Assessing a modified-Sonine kinetic theory for instabilities in highly dissipative, cooling granular gases,'' Phys. Rev. E \textbf{85}, 041303 (2012).

\bibitem{G05}V. Garz\'o, ``Instabilities in a free granular fluid described by the Enskog equation,'' Phys. Rev. E \textbf{72}, 021106 (2005).

\bibitem{exp1}M. G. Clerc, P. Cordero, J. Dunstan, K. Huff, N. M\'ujica, D. Risso and G. Varas, ``Liquid-solid-like transition in quasi-one-dimensional driven granular media,'' Nature Phys.
{\bf 4}, 249 (2008); F. Vega Reyes and J. S. Urbach, ``Effect of inelasticity on the phase transitions of a thin vibrated granular layer,'' Phys. Rev. E {\bf 78}, 051301 (2008).


\bibitem{DHDNZ10}C. M. Donahue, C. M. Hrenya, R. H. Davis, K. J. Nakagawa, A. P. Zelinskaya, and G. G. Joseph, ``StokesÆ cradle: normal three-body collisions between wetted particles,'' J. Fluid Mech. \textbf{650}, 479 (2010).




\bibitem{GFM09}V. Garz\'o, F. Vega Reyes and J. M. Montanero, ``Modified Sonine approximation for granular binary mixtures,'' J. Fluid Mech. {\bf 623}, 387 (2009).

\bibitem{G03}I. Goldhirsch, ``Rapid granular flows,'' Annu. Rev. Fluid Mech. \textbf{35}, 267 (2003).

\bibitem{SGD04}A. Santos, V. Garz\'o and J. W. Dufty,  ``Inherent rheology of a granular fluid in uniform shear flow,'' Phys. Rev. E \textbf{69}, 061303 (2004).




\bibitem{VU09}F. {Vega Reyes} and J. S. Urbach, ``Steady base states for Navier-Stokes granular hydrodynamics with boundary heating and shear,'' J. Fluid Mech. \textbf{636}, 279 (2009).

\bibitem{VSG13}F. {Vega Reyes}, A. Santos and V. Garz\'o, ``Steady base states for non-Newtonian granular hydrodynamics,'' J. Fluid Mech. \textbf{719}, 431 (2013).

\bibitem{VSG10}F. {Vega Reyes}, A. Santos and V. Garz\'o, ``Non-Newtonian granular hydrodynamics. What do the inelastic simple shear flow and the elastic Fourier flow have in common?,'' Phys. Rev. Lett. \textbf{104}, 028001 (2010).


\bibitem{GT96}I. Goldhirsch and M-L. Tan, ``The single particle distribution function for rapid granular shear flows of smooth inelastic disks,'' Phys. Fluids \textbf{8}, 1752 (1996).

\bibitem{SG98}N. Sela and I. Goldhirsch, ``Hydrodynamic equations for rapid flows of smooth inelastic spheres to Burnett order,'' J. Fluid Mech. \textbf{361}, 41 (1996).



\bibitem{Brey2}J. J. Brey, M. J. Ruiz-Montero, and F. Moreno, ``Hydrodynamics of an open vibrated granular system,'' Phys. Rev. E, \textbf{63}, 061305 (2001); J. J. Brey, M. J. Ruiz-Montero, F. Moreno, and and R. Garc\'{\i}a-Rojo, ``Transversal inhomogeneities in dilute vibrofluidized granular fluids,''  \emph{ibid.}, \textbf{65}, 061302 (2002).


\bibitem{L02}J. Lutsko, J. J. Brey, and J. W. Dufty, ``Diffusion in a garanular fluid. Part 2. Simulation,'' Phys. Rev. E \textbf{65}, 051304 (2002).

\bibitem{LLC07}G. Lois, A. Lema\^{\i}tre, and J. M. Carlson, ``Spatial force correlations in granular shear flow. Part 2. Theoretical implications,'' Phys. Rev. E \textbf{76}, 021303 (2007).

\bibitem{BGGL09}M. N. Bannerman, T. E. Green, P. Grassia, and L. Lue, ``Collision statistics in sheared inelastic hard spheres,'' Phys. Rev. E \textbf{79}, 041308 (2009).

\bibitem{R02}E. C. Rericha, C. Bizon, M. D. Shattuck, and H. L. Swinney, ``Shocks in supersonic sand,'' Phys. Rev. Lett. \textbf{88}, 014 302 (2002).

\bibitem{Y02}X. Yang, C. Huan, D. Candela, R. W. Mair, and R. L. Walsworth, ``Measurements of grain motion in a dense, three-dimensional granular fluid,'' Phys. Rev. Lett. \textbf{88},044301 (2002); C. Huan, X. Yang, D. Candela, R. W. Mair, and R. L. Walsworth, ``NMR experiments on a three-dimensional vibrofluidized granular medium,'' Phys. Rev. E \textbf{69}, 041302 (2004).

\bibitem{FK}J. Ferziger and H. Kaper, {\em Mathematical Theory of Transport Processes in Gases}
(North-Holland, Amsterdam, 1972).

\bibitem{GD99a}  V. Garz\'{o} and J. W. Dufty, ``Homogeneous cooling
state for a granular mixture,'' Phys. Rev. E {\bf 60}, 5706 (1999).

\bibitem{BRCG00} J. J. Brey, M. J. Ruiz-Montero, D. Cubero, and R. Garc\'{\i}a-Rojo, ``Self-diffusion in
freely evolving granular gases,'' Phys. Fluids {\bf 12}, 876 (2000).

\bibitem{footnote}See supplementary material at [URL will be inserted by AIP] for a \emph{Mathematica} code that evaluates the diffusion transport coefficients in the first and second Sonine approximations for arbitrary values of composition, masses, diameters and coefficients of restitution. This code can be also download from \url{http://www.eweb.unex.es/fisteor/vicente/granular_files.html}.





\bibitem{MC84}M. L\'opez de Haro and E. G. D. Cohen, ``The Enskog theory for multicomponent mixtures: III. Transport properties of dense binary mixtures with one tracer component,'' J. Chem. Phys. \textbf{80}, 408 (1984).

\bibitem{Lambda}K. E. Grew and T. L. Ibbs, \emph{Thermal Diffusion in Gases} (Cambridge University Press, Cambridge, 1952); G. C. Maitland, M. Rigby, E. B. Smith, and W. A. Wakeman, \emph{Intermolecular Forces: Their Origin and Determination} (Clarendon, Oxford, 1981).

\bibitem{G06}V. Garz\'o, ``Segregation in granular binary mixtures: Thermal diffusion,'' Europhys. Lett. {\bf 75}, 521 (2006).

\bibitem{G08}V. Garz\'o, ``Brazil-nut effect versus reverse Brazil-nut effect in a moderately granular dense gas,'' Phys. Rev. E {\bf 78}, 020301(R) (2008); ``Segregation by thermal diffusion in moderately dense granular mixtures,'' Eur. Phys. J. E \textbf{29}, 261(2009).

\bibitem{G11}V. Garz\'o, ``Thermal diffusion segregation in granular binary mixtures described by the Enskog equation,'' New J. Phys. \textbf{13}, 055020 (2011).

\bibitem{BRM05}J. J. Brey, M. J. Ruiz-Montero, and F. Moreno, ``Energy partition
and segregation for an intruder in a vibrated granular system
under gravity,'' Phys. Rev. Lett. {\bf 95}, 098001 (2005); ``Hydrodynamic profiles for an impurity in an open vibrated granular gas,'' Phys. Rev. E {\bf 73}, 031301 (2006).


\bibitem{BKD11}J. J. Brey, N. Khalil, and J. W. Dufty, ``Thermal segregation beyond Navier-Stokes,'' New J. Phys. \textbf{13}, 055019 (2011); ``Thermal segregation of intruders in the Fourier state of a granular gas,'' Phys. Rev. E \textbf{85}, 021307 (2012).

\bibitem{BT02}A. Barrat and E. Trizac, ``Molecular dynamics simulations of vibrated granular gases,'' Phys. Rev. E \textbf{66}, 051303 (2002).

\bibitem{BEGS08}R. Brito, H. Enr\'{\i}quez, S. Godoy, and R. Soto, ``Segregation induced by inelasticity in a vibrofluidized granular mixture,'' Phys. Rev. E \textbf{77}, 061301 (2008); R. Brito and R. Soto, ``Competition of Brazil nut effect, buoyancy, and inelasticity induced segregation in a granular mixture,'' Eur. Phys. J. Special Topics \textbf{179}, 207 (2009).





\end{thebibliography}
\end{document}